\documentclass{aa}
\usepackage{times}
\usepackage{journal}
\usepackage{graphics}
\usepackage{aabib99}
\begin{document}

\thesaurus{11(11.05.2; 13.21.1; 13.09.1; 13.18.1)}

\title{Galaxy Modelling --- I. Spectral Energy Distributions
from Far--UV to Sub--mm Wavelengths}

\author{ Julien E.\,G.\,Devriendt\inst{1}
         \and Bruno Guiderdoni\inst{1} 
         \and  Rachida Sadat\inst{1,2,3}} 

\institute{ Institut d'Astrophysique de Paris,
            98\,{\it bis}, boulevard Arago, F75014 Paris, France 
          \and Observatoire Astronomique, 
            11, rue de l'Universit\'e, F67000 Strasbourg, France 
          \and C.R.A.A.G., BP 63, Bouzareah, Algiers, Algeria }

\offprints{\tt devriend@iap.fr}

\authorrunning{J. Devriendt et al.}
\titlerunning{I. Spectral Energy Distributions of Galaxies}

\date{Received 1999 April 23  / Accepted 1999 June ...}

\maketitle

\begin{abstract}

We present {\sc STARDUST}, a new self--consistent modelling of the spectral
energy distributions (SEDs) of galaxies from far--UV to
radio wavelengths. In order to derive the SEDs in this broad spectral range, 
we first couple spectrophotometric and (closed--box) chemical evolutions 
to account for metallicity effects on the spectra of synthetic stellar
populations. We briefly compare the UV/visible/near--IR colours and 
magnitudes predicted by our code with those of other codes available in the 
literature and we find an overall agreement, in spite of 
differences in the stellar data. We then use a phenomenological fit for 
the metal--dependent extinction curve and a simple geometric distribution 
of the dust to compute the optical depth of galaxies and the corresponding 
obscuration curve. This enables us to 
calculate the fraction of stellar light reprocessed in the infrared range. 
In a final step, we define a dust model with various components and we fix 
the weights of these components in order to reproduce the IRAS correlation of 
IR colours with total IR luminosities. This allows us to compute far--IR SEDs
that phenomenologically mimic observed trends.
We are able to predict the spectral evolution of galaxies in a broad 
wavelength range, and we can reproduce the observed SEDs of local spirals, 
starbursts, luminous infrared galaxies (LIRGs) and ultra luminous 
infrared galaxies (ULIRGs). This modelling is so far kept as simple as 
possible and depends on a small number of free parameters, namely the 
initial mass function (IMF), star formation rate 
(SFR) time scale, gas density, and galaxy age, as well as on more refined 
assumptions on dust properties and the presence (or absence) of gas 
inflows/outflows. However, these SEDs will be subsequently
implemented in a semi--analytic approach of galaxy formation, where most of
the free parameters can be consistently computed from more general 
assumptions for the physical processes ruling galaxy formation and evolution.


\end{abstract}

\section{Introduction}

Our knowledge of galaxies is essentially based upon the
light they emit, so that anyone interested in studying
such objects has to address the obvious --- but nevertheless
complex --- issue of their spectrophotometric evolution.
This becomes even more difficult if one is interested in 
handling their luminosity budget properly. Part of the light emitted
by stars is reprocessed in the mid/far--IR window by the dust
these very same stars produce, so that our knowledge of the star 
formation history of an individual galaxy (and a fortiori
the overall star formation history of the universe) crucially depends on our 
ability to estimate the shape of its spectrum from the far--UV to 
the submm wavelength range.  The purpose of this paper is to compute 
SEDs in the broadest wavelength range, that can be used to analyze the 
increasing amount of multiwavelength observations targetting high-z 
galaxies. 

As a matter of fact, several pieces of evidence are now converging 
to give a novel view of forming galaxies' luminosity budget. 
This can be summarized as follows.

The discovery of the cosmic infrared background (CIRB) by 
the COBE satellite, is the first
evidence that high-z galaxies strongly 
emit in the IR/submm range, and that dust played an important role in the 
luminosity budget of these galaxies (Puget et al. \cite*{P+al96};
Guiderdoni et al. \cite*{GBPLH97};
Fixsen et al. \cite*{FDMBS98}, Schlegel et al. \cite*{SFD98}, Hauser et
al. \cite*{H+al98}). Though the exact 
level of the CIRB is still a matter of debate \cite{LABDP99}, it appears to be 
5-10 times higher than the no-evolution prediction based on the IRAS local luminosity 
function, and 2 times more intense than the cosmic optical background \cite{PMZFB98}.

At 850 $\mu$m, the background has been broken into its components by
SCUBA at the JCMT 
(Smail et al. \cite*{SIB97}, Hughes et al. \cite*{HS+al98}, Barger et
al. \cite*{BCSFTSKO98}, Eales et al. \cite*{ELGDBHLC98}). 
The same work has been achieved respectively by ISOPHOT 
at 175 $\mu$m (Kawara et al. \cite*{K+al98}, Puget et al. \cite*{P+al99}, Dole et al. \cite*{D+al99}), and ISOCAM  
at 15 $\mu$m (Oliver et al. \cite*{O+al97}, Elbaz et
al. \cite*{E+al98},\cite*{E+al99}), both on board the ISO satellite. 
Although getting theoretical identifications
and redshifts of such objects is not an easy task (Lilly et al. \cite*{L+al99}, 
Barger et al. \cite*{BCSIBK99}), the sources seem to be the high--redshift analogs of the
local ``luminous'' and ``ultraluminous infrared galaxies'' (LIRGs and ULIRGs).
However, there are uncertainties on the properties of high-redshift dust, 
and on the mechanism that is responsible for dust heating (starbursts versus
active galactic nuclei). In local LIRGs and ULIRGs samples, starbursts dominate, except in the brightest 
sources, with $ L > 3 \times 10^{12} L_\odot$ (Genzel et al. \cite*{G+al98}, Lutz et al. \cite*{LSRMG98}). We will 
hereafter assume that starbursts are the component that dominates dust heating
and we will neglect the influence of the AGNs.

Optical studies are also showing that high-z 
galaxies selected from their optical properties, and in particular the LBGs
at $z = $ 3 and 4 (Steidel et al. \cite*{SGPDA96}, Madau et al. \cite*{MF+al96}), are so heavily 
extinguished that probing their properties requires multi-wavelength 
observations. Current estimates (Flores et al. \cite*{FH+al99},
Steidel et al. \cite*{SAGDP99}, Meurer et al. \cite*{MHC99}) give
extinction 
factors $0.1 < E (B-V) < 0.4$, with a large scatter 
and the trend that the brightest objects are also the more
extinguished ones. 
The net effect is that the SFR estimated from UV fluxes are too low by a 
factor 3 -- 5 on average.
Observations seem to show that dust is present at redshifts beyond 4 since 
it is seen, for instance, by its extinction of a gravitationally lensed 
galaxy at z=4.92 (Soifer et al. \cite*{SNFMI98}) and by its mm emission in radio galaxies 
and quasars (e.g. BR1202-0725 at z=4.69, McMahon et al. \cite*{MOBKH94}). 

So there is clearly a need for models of SEDs that 
take the extinction and emission effects of dust into account in a consistent 
way. Since the pioneering work by Tinsley \cite*{T72}, 
Searle et al. \cite*{SSB73} and Huchra \cite*{H77b},
 where the photometric evolution was restricted
to the visible wavelength range, stellar population 
synthesis models were designed \cite{B83} and extended
to include the nebular component and extinction \cite{GRV87}.
These models were driven by the need to reproduce observed 
 spectrophotometric properties of local as well as more distant 
galaxies from the far--UV to the near--IR wavelength range.   
Spectrophotometric evolution models were then improved 
(Bruzual and Charlot, 1993; Fioc and Rocca--Volmerange, 1997) 
to include tracks with various metallicities, and late stages of stellar
evolution.
They also gained a better time resolution thanks to the isochrone scheme 
\cite{CB91}. Work was also  
achieved to couple chemical and spectrophotometric models
self--consistently \cite{A89}.

Other models were designed, where, in addition to chemical evolution, the photometric 
evolution of stellar populations which heat the dust is taken
into account (Franceschini et al. \cite*{FZTMD91}, \cite*{FMZD94} 
and Fall, Charlot and Pei \cite*{FCP96}, Silva et al. \cite*{SGBD98}). They all assume
a simple relationship between the dust content and the heavy--element 
abundance of the gas (for details on the complex processes of dust
grain creation and destruction, see e.g. Dwek \cite*{D98} and 
references therein) and do not have a detailed modelling of
the re-processing of starlight by dust grains. 
On the other hand, models restricted to galaxy emission at IR/submm wavelengths 
have been developed (see eg. Rowan-Robinson and Crawford 
\cite*{RRC89}, Maffei \cite*{Mphd94}, Guiderdoni et al. 
\cite*{GHBM98}) but none of them describe in a 
self--consistent and detailed way the stellar source that powers 
the emission. We would like to emphasize the need for 
models which explicitly connect both UV/optical and IR/submm windows, 
because they are the only ones capable of correctly 
handling the luminosity budget of galaxies and hence their star
formation histories.
As a matter of fact, these two windows 
are {\em naturally} linked together. 
For it is dust, which partially absorbs starlight 
coming out in the UV/optical window and re--emits it in 
the far--IR. 

In this paper, we present such a model, {\sc STARDUST},
which contains up--to--date spectrophotometric modelling
self--consistently coupled to chemical evolution and dust
absorption (following the same guideline as Guiderdoni and Rocca--Volmerange \cite*{GRV87},
Franceschini et al. \cite*{FMZD94} and Guiderdoni et
al. \cite*{GHBM98}). We further
describe a new dust emission model following the lines of 
a model developed by Maffei \cite*{Mphd94}. Our approach 
differs from the one of Silva et al. \cite*{SGBD98} in the sense that
we empirically adjust our SEDs in the IR in order
to match average IRAS colors, instead of trying to derive them from basic 
physical arguments. What is lost in the understanding of the 
complicated physical processes involving dust is gained 
in the smaller number of parameters required by the model.  
Bearing this in mind, the model allows one to make consistent predictions 
for the SEDs of galaxies in a very broad wavelength range.

More specifically, most of the free parameters necessary to obtain such SEDs
can be self--consistently computed in the explicit cosmological framework of semi--analytic
models of galaxy formation and evolution (SAMs). However, we defer to
a companion paper \cite{DG99} a discussion
of the results obtained by implementing our SEDs in such SAMs. 
In Section 2, we briefly describe the spectro-photometric model.
Section 3 deals with chemical evolution, Section 4 with the dust.
In Section 5 we put the pieces together to build the complete
spectral energy distribution of the galaxies. Section 6  
sketches out what a ``dusty'' high--redshift universe 
might look like, and we eventually draw conclusions in Section 7.

\section{Computing Spectra of the Stellar Component}

\subsection{Numerical Scheme}

The goal is to compute the stellar $F^*_\lambda (t)$ flux
contribution to the galactic spectrum at time $t$. 
Such a flux can be written as follows:
\begin{equation}
F^*_\lambda (t)= \int^t_0 \int^{m_u}_{m_d} \psi_*(t-\tau)
\phi(m) f_\lambda(m,\tau) {\rm d}m \, {\rm d}\tau  \, \, ,
\end{equation} 
where $m_d$ is the minimal star mass, $m_u$ is the 
upper mass cutoff (respectively $0.1 M_\odot$, and 
$120 M_\odot$ here),
$\psi_*(t-\tau)$ is the number of $M_\odot$ of gas that gets 
turned into stars at time $(t-\tau)$ per time unit, 
$\phi(m) \propto m^{-x}$ (in the following we take 
$x=1.35$ for $m$ in $[m_d,m_u]$ \cite{S55}) is 
the IMF normalized to 1 $M_\odot$,
 and $f_\lambda(m,\tau)$ is the flux of a star of 
initial mass $m$ at wavelength $\lambda$ and age $\tau$
($\tau = 0$ corresponds to the zero age main sequence, and
$f_\lambda(m,\tau) = 0$ if $\tau > t_*(m)$ where $t_*(m)$ is
the life time of a star of mass $m$).  
We shall neglect the nebular component in the following.

The straightforward solution consists in discretizing this double
integral, 
but leads to oscillations in the resulting flux 
mainly due to the rapidly evolving stellar 
stages \cite{CB91}.
The natural method to get around this, the so--called
``isomass scheme'' only discretizes the
integral over mass \cite{GRV87}. However, the problem
with such an approach is that it requires 
a very fine mass grid in order to achieve a good overlap
of equivalent evolutionary stages for consecutive masses.
Therefore, it is computer time consuming \cite{Fphd97}. 
The other solution that leads to the same results
but is much more computationally tractable is called
the ``isochrone scheme'' and is based upon the  
discretization of the integral over time \cite{CB91}. It
has the advantage of being directly comparable to 
color--magnitude diagrams of star clusters.
The isochrones are of course carefully picked in order not
to miss any stage of stellar evolution.
We use such a scheme in the model presented here.  

\begin{figure}[htbp]

\resizebox{9cm}{!}{\includegraphics{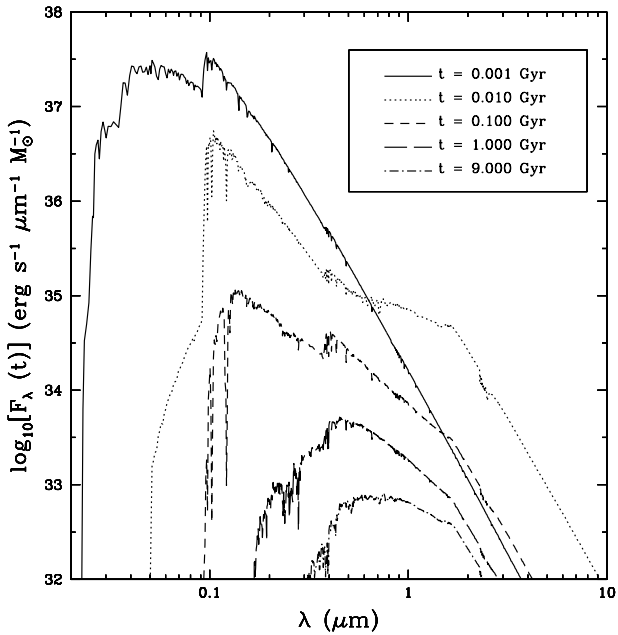}}
\hfill
\parbox[b]{87.5mm} {\caption{Spectral evolution
of an instantaneous burst of star formation with solar
metallicity.
The IMF is Salpeter, and the flux level corresponds
to a $1~M_\odot$ galaxy.}
  \label{ps_evol_Zsol}}

\end{figure}

\subsection{Library of Stellar Tracks} 

Our model is mainly based on the so-called ``Geneva tracks'' 
(Schaller et al. \cite*{SSMM92}, Schaerer et al. \cite*{SCMMS93}
, Charbonnel et al. \cite*{CMMSS93}, Schaerer et al. \cite*{SMMS93}) 
($Z =$ 0.001, 0.004, 0.008, 0.02, 0.04 and 
0.8 $\le M/M_{\odot} \le$ 120 ). For stars less massive
than $1.7 M_{\odot}$, these tracks stop at 
the Giant Branch tip. 
Hence more recent grids of models, based on Geneva tracks and 
covering the evolution of low mass stars 
(0.8 to 1.7 $M_{\odot}$) from the zero age main sequence 
up to the end of the early-AGB, are included for $Z=$ 0.001 
and $Z=$ 0.02 \cite{CMMS96}.
For the late stages of other metallicities (horizontal branch,
earlt AGB), we either interpolate or 
extrapolate $\log L_\mathrm{bol}$, $\log T_\mathrm{eff}$ 
and $\log t$ versus $\log Z$ from the available tracks 
(Meynet private communication).  
The final stages of the stellar evolution (thermally 
pulsing-AGB and post-AGB) are not included.

The main motivation for choosing the Geneva Group tracks 
library is the possibility to use $Z$-dependent stellar 
yields (Maeder \cite*{M92}, Maeder \cite*{M93}) which insures 
consistency when spectrophotometric and chemical 
evolutions are coupled. The initial helium content of the gas from which 
stars form is computed starting with
an initial helium fraction $Y=$ 0.24 after primordial
nucleosynthesis and assuming an enrichment rate compared to metals  
of $\frac{\Delta Y}{\Delta Z}=3$ (so that $Y=$ 0.3 for solar
metallicity).

\subsection{Stellar Spectra}

An obvious advantage of using theoretical instead of 
empirical stellar spectra, is that the 3-parameter space
of the theoretical spectra ($\log T_\mathrm{eff}$,
 $\log L_\mathrm{bol}$ (or $\log R$) and $\log g$), is similar to 
the one of the HR diagram. This avoids the use of the 
transformations of bolometric luminosity $\log L_\mathrm{bol}$ 
to visual magnitude $M_{V}$, and of $\log T_\mathrm{eff}$ 
to spectral types and luminosity 
classes, which are rather uncertain for the hottest and 
coldest stars.

Here we use the theoretical fluxes grid of Kurucz \cite*{KU92} 
(hereafter K92) which covers all metallicities 
from $\log Z/Z_\odot = +1.0 $ to $\log Z/Z_\odot =-5.0 $ and 61 temperatures from 3500 K to 
50 000 K. Each spectrum spans a wavelength range between 
90 $\rm \AA$ and 160 $\mu$m, with a mean resolution of 
20 $\rm \AA$. 
For the coldest stars (K and M-type stars) with $T \le$ 3750 K,
 K92 models fail to reproduce the observed spectra.
Therefore, we prefer to use models coming from 
different sources. For M--giants, the Bessel et al. 
\cite*{BBWS89},\cite*{BBSW91a},\cite*{BBSW91b}, 
\cite*{BWBS91} (hereafter BBSW) models have been used. 
The wavelength range of BBSW fluxes is restricted to 
0.49--5 $\mu$m. We have added a 
blackbody tail to the red end part of the spectra. 
On the other hand, BBSW M--giants present a spurious 
spike around 5000 $\rm \AA$, which we correct 
for, using the following procedure:
we replace the BBSW fluxes downward of 0.6 
$\mu$m by a handful of Gunn and Stryker \cite*{GS83} M giants 
fluxes. To do this, we recompute 
the Gunn and Stryker spectra of known spectral type 
and V--K colour for the effective temperature of BBSW 
spectra. Then we add  
a blackbody tail redward of 0.7 $\mu$m  \cite{W94}. 
As noted by this author, Gunn and Stryker fluxes are not 
suitable downward from 0.36 $\mu$m. We thus attach 
a blackbody for shorter wavelengths. 
In a similar manner, the M dwarf sequence of BBSW is 
adopted (with $ \log g=$ 4.5--5) but instead of adding Gunn and
Stryker spectra at the position of the spike, we
use M dwarf models computed by Brett 
\cite*{B95a} \cite*{B95b}. Blackbody tails are 
added to complete the flux library.

For both BBSW and Brett models, we rebin the 
spectra using the same wavelength grid as in K92. 
Furthermore, BBSW and K92 grids cover overlapping 
metallicity ranges which we interpolate to the 
five metallicities imposed by the Geneva tracks.

\subsection{Post--Starburst Evolution}

With the material that has been described in the previous 
sections, one can build the spectrum of what is called a simple 
stellar population (SSP).
This simply means that at time $t$=0, one
turns all the gas ($1 M_\odot$ here) into stars which
are distributed over the mass range $[m_d,m_u]$ according
to the Salpeter IMF and one lets them evolve passively.  
Such a time evolution is shown in Fig. \ref{ps_evol_Zsol}. 

\begin{figure}[htbp]

\resizebox{9cm}{!}{\includegraphics{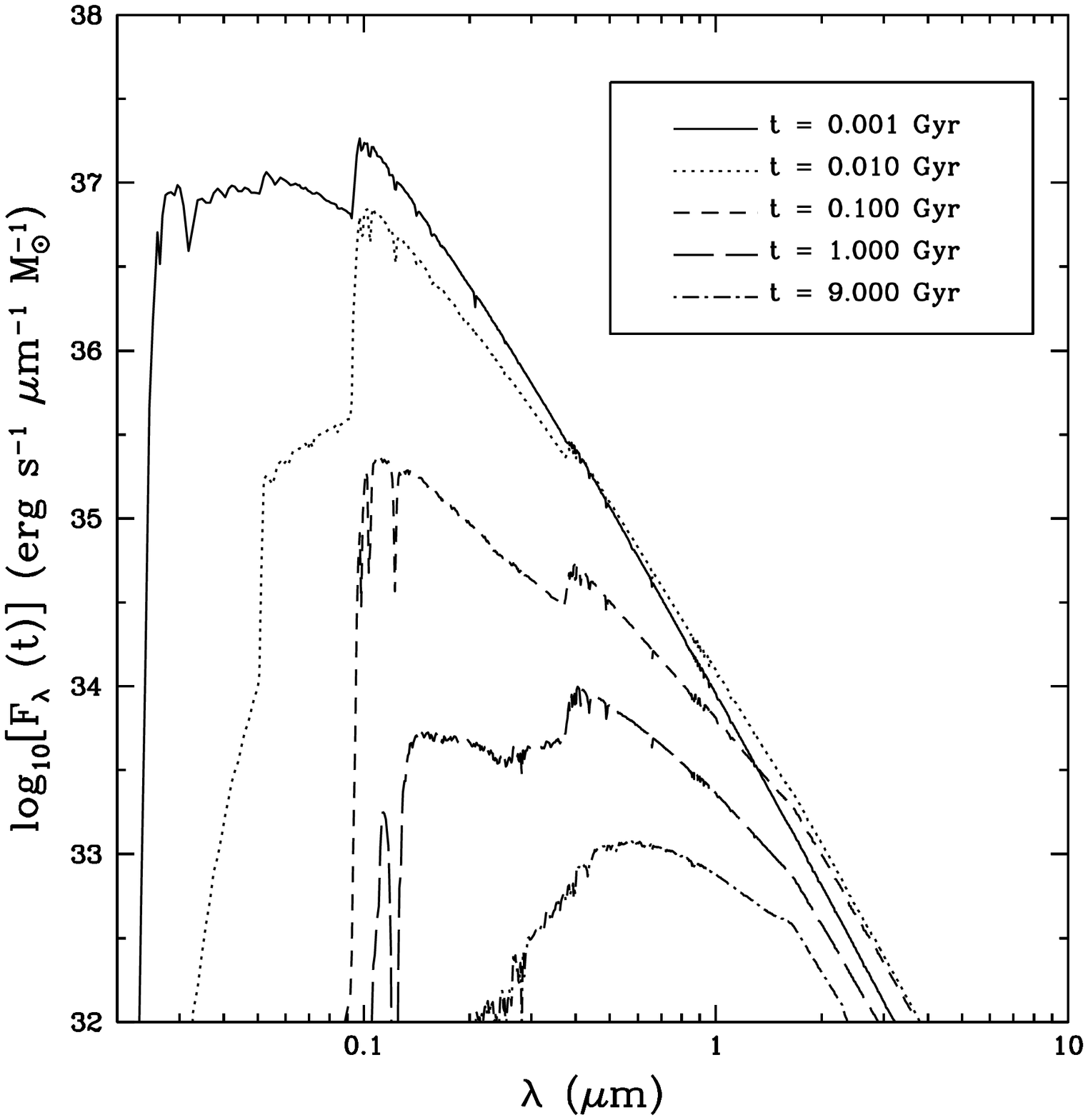}}
\hfill
\parbox[b]{87.5mm} {\caption{Same as 
Fig.~\ref{ps_evol_Zsol} but for a ${1 \over 20} Z_\odot$ 
burst.}
  \label{ps_evol_Zlow}}

\end{figure}

What is noticeable at first sight is the huge decrease
of the flux with time which is all the more important when the 
wavelength is short, and simply reflects the fact that the most
luminous stars are the most massive ones and also the first ones to die.
The second point is that one can clearly see the red super giant stars
(RSG) population building up in the infrared longward of 1 $\mu$m
between 0.01 and 0.06 Gyr.

These RSG stars are very sensitive to the metallicity of the initial gas
from which they were born, in the
sense that the higher the metallicity of this gas, the higher the 
flux emitted in the near-IR by the RSGs. This is clearly illustrated
in Fig.~\ref{ps_evol_Zlow} for which the stars formed from a metal--
poor gas. At $t =$ 0.01 Gyr, the flux at wavelengths
greater than 1 $\mu$m is about an order of magnitude lower than
the flux for an identical population of stars that formed from initial
gas with solar metallicity (Fig.~\ref{ps_evol_Zsol}).
Comparing the two figures also allows one to check
the well--known trend that the higher the metallicity of the initial
gas, the fainter the B band magnitude for $t \geq 1$ Gyr.

\begin{figure}[htbp]

\resizebox{9cm}{!}{\includegraphics{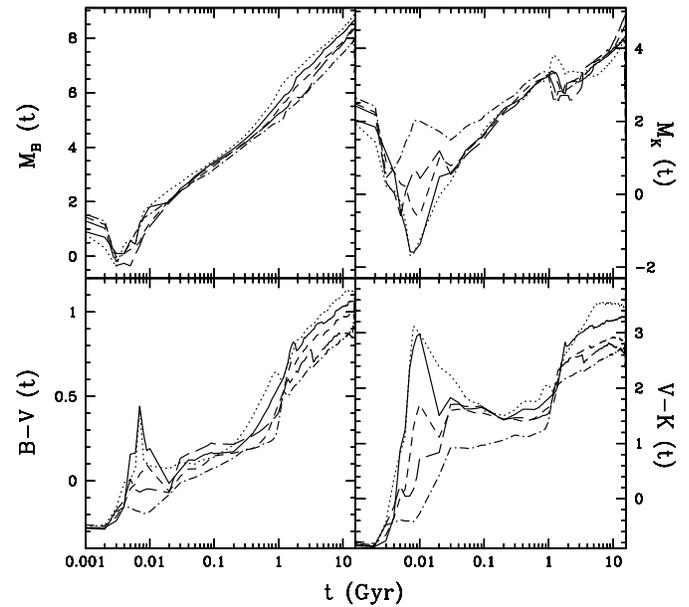}}
\hfill
\parbox[b]{87.5mm} {\caption{Magnitude and color evolution
of an instantaneous burst with a Salpeter IMF. The solid
 curve corresponds to solar, the dotted line to
twice solar, short--dashes to 2/5 solar, long-dashes to
1/5 solar and the dot--dashed line to 1/20 solar metallicities.
Magnitudes are normalized to $1~M_\odot$.}
  \label{ps_lumcol}}

\end{figure}

If order to compare such spectra directly with
color data, one just has to convolve the SED
with filters used to observe real galaxies. 
Color and magnitude
 evolutions are obtained for instantaneous bursts of star 
formation. This is shown in Fig. \ref{ps_lumcol}.
A typical application for such a 
model is the study of globular clusters.

\subsection{Comparing Codes}

\begin{figure}[htbp]
\resizebox{9cm}{!}{\includegraphics{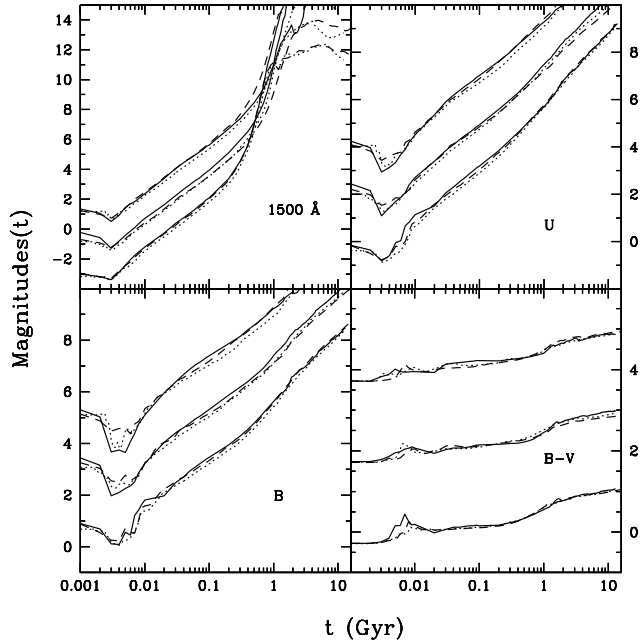}}
\hfill
\parbox[b]{87.5mm}{\caption{Comparison of {\sc GISSEL} (version 
1998) (dotted line) (Bruzual and Charlot, 1993), {\sc P\'EGASE} 
(dashed line) (Fioc and Rocca--Volmerange, 1997), and our 
{\sc STARDUST} (solid line) for an instantaneous burst with solar 
metallicity (lowest curve), 2/5 solar (middle curve),
and 1/5 solar (upper curve) in the different wavelength bands 
indicated on the panels. For a better visualization the 
1/5 solar and 2/5 solar metallicities have been
shifted by 2 and 4 magnitudes respectively.}
\label{mcomp_opt}}

\end{figure}

\begin{figure}[htbp]
\resizebox{9cm}{!}{\includegraphics{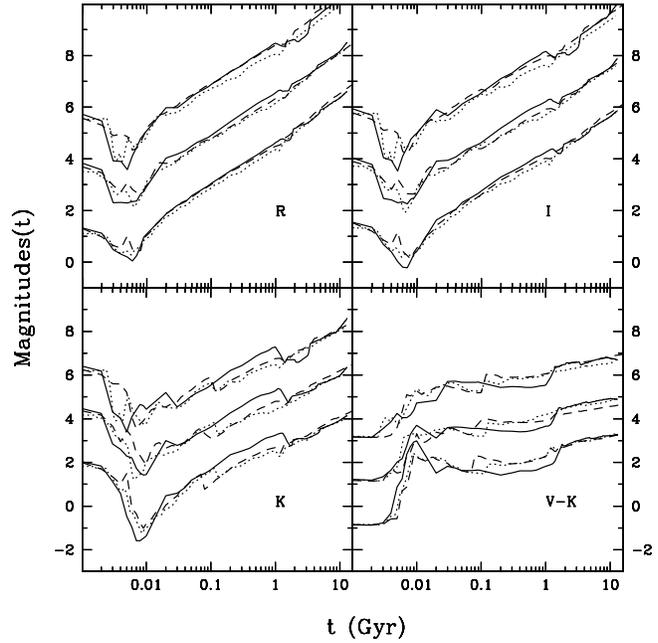}}
\hfill
\parbox[b]{87.5mm}{\caption{Same as Fig.~\ref{mcomp_opt}.}
\label{mcomp_ir}}

\end{figure}

The next interesting step is to see how our code 
compares to other population synthesis codes.
Readers interested in more detailed comparisons 
(although on older versions of similar codes) are referred 
to Charlot et al. \cite*{CWB96} and the reviews by Arimoto \cite*{A96} 
and Charlot \cite*{C96}.
In Fig.\ref{mcomp_opt} and 
\ref{mcomp_ir}, we show a
comparison of the photometry predicted by three 
different codes for three different metallicities. 
For the three codes, a similar standard Salpeter IMF has been used to
distribute the stars in 
an instantaneous burst of star formation that takes place at $t$=0.

The P\'EGASE code \cite{Fphd97} uses the Padova tracks \cite{BFBC93} 
for stellar
evolution, with an initial helium fraction $Y=$ 0.23 and an enrichment
of $\frac{\Delta Y}{\Delta Z}=2.5$, up to the end of the early AGB
phase. It also includes the thermally-- 
pulsing AGB \cite{GdJ93}, and the post--AGB stages from Sch\"onberner \cite*{S83}
and Bl\"ocker \cite*{B95}. The spectral libraries are from Lejeune et al. \cite*{LCB97}. 
The GISSEL code (version 1998) uses tracks from Padova up to the early AGB phase
which are completed till the end of the white dwarf cooling sequence
and includes the thermally--pulsing, the post--AGB stages, as well as 
carbon stars (Charlot, private communication). The spectral libraries are the same 
as for P\'EGASE.

As can be seen in the different panels of Fig.~\ref{mcomp_opt} and
\ref{mcomp_ir}, models give very similar results in all bands, from
the far--UV to the near--IR, although they use different stellar tracks
and spectra. Most of the time, the differences are  
minor (around 0.1 magnitude), especially for colors and metallicities 
close to the solar value. However there are some
discrepancies for early stages of evolution
($t \le 3 $Myr), in the far--UV for $t \ge 1 $Gyr, and in
the near--IR (mainly the K band).

As far as the early evolution is concerned, although we are in fair agreement
(better than 0.3 magnitudes) with the P\'EGASE and GISSEL models for all
metallicities and wavebands, the uncertainties in the 
early evolutionary stages of massive stars (essentially mass loss) are responsible 
for the larger flux difference between the models.

In the far--UV, after about 1 Gyr,  
differences are due to the uncertainties in the contribution 
of intermediate to low mass stars to the flux at short wavelengths. This
is not relevant as long as one does not try to analyze E/SO 
galaxy spectra in this wavelength range, because, as mentioned e.g. by Rocca--Volmerange
and Guiderdoni \cite*{RVG87}, the level of such a flux is 
negligible compared to the far--UV flux
produced by even a small amount of star formation.

In the near--IR, the discrepancies (more pronounced for
low metallicity bursts where they can attain 0.5
magnitude ) are due to the
modelling of the final stages of stellar evolution (thermally--
pulsing AGB and post--AGB) that are taken into account
in P\'EGASE and GISSEL but not in our model. The thermally--
pulsing AGB stars are responsible for the excess of flux of
P\'EGASE and GISSEL in the near--IR (K band) and the redder V-K color from 
about 100 Myr to 1 Gyr.

We want to emphasize the point that the 
comparison we have
done here is by no means extensive, and is likely to depend
on the IMF that one picks. Nevertheless, the remarkably
good agreement observed in this specific case 
between the three totally independent codes is fairly 
encouraging.

\section{Chemical Evolution}

As previously mentioned, we use the Z-dependent stellar 
yields of Maeder \cite*{M92} with moderate mass loss for high mass 
stars. For intermediate mass stars, we use yields given by 
Renzini and Voli \cite*{RV81}. We also include type Ia supernova whose modelling
and yields respectively follow Ferrini et al. \cite*{FMPP92} and Nomoto et al. 
\cite*{NTYH95}.

Our chemical evolution model is a simple closed box model
(see e.g. Tinsley \cite*{T72}),
 where we assume that the sum of the mass of stars $M_*(t)$ and
 gas $M_g(t)$ remains constant through time:
\begin{equation} 
M_g (t) + M_* (t) = M_{tot}~,~ \, M_{tot}=\mathrm{C^{st}}
\label{eqgas}
\end{equation}

We do not make the assumption of instantaneous recycling 
but we track metals ejected by stars at each time 
step and let new stars form out of the enriched gas, 
assuming instantaneous mixing of the metals. In other 
words, we solve at each time step the following equation :
\begin{equation}
\frac{d(Z_g M_g)(t)}{dt}=-Z_g(t)\psi_*(t) 
+ \mathcal{E}_Z(t)
\label{eqmet}
\end{equation}
where $Z_g$ is the mass percentage of element Z in the gas,
$\psi_*(t)$ is the amount of gas consumed by star 
formation, {\em i.e.} the SFR
and is taken to be:
\begin{equation}
\psi_*(t) = \frac{M_g(t)}{t_*}
\label{eqsfr}
\end{equation} 
where $t_*$ is referred to as the characteristic time scale 
for star formation.

Finally, $\mathcal{E}_Z(t)$ is the ejection rate of element Z
at time t:
\begin{eqnarray}
\mathcal{E}_Z(t) & = & \int_{m_t}^{m_u} d(\log m) 
\psi_*(t-t(m)) \phi(m)
\nonumber \\ 
& \times  & \left\{ (m-w(m)) Z(t-t(m)) + m Y_Z(m) \right\} ,
\end{eqnarray}
where $m_t$ is the minimum possible star mass ( i.e. the mass 
of a star having a lifetime of $t$).

The first term between curly braces, $(m-w(m))Z(t-t_m)$, is the ejected 
mass of initially present metals ( i.e. the difference between the initial mass of the star 
and the remnant mass, times the initial metallicity), the $(t-t(m))$ 
argument being due to the fact that stars of mass $m$ which 
die at time $t$ were born at $(t-t(m))$.
The second term in these braces, $m Y_Z(m)$, is the 
percentage of the initial star mass transformed into new 
elements Z (i.e. the stellar nucleosynthesis). $Y_Z(m)$ 
is referred to as the stellar yield.

\begin{figure}[htbp]

\resizebox{9cm}{!}{\includegraphics{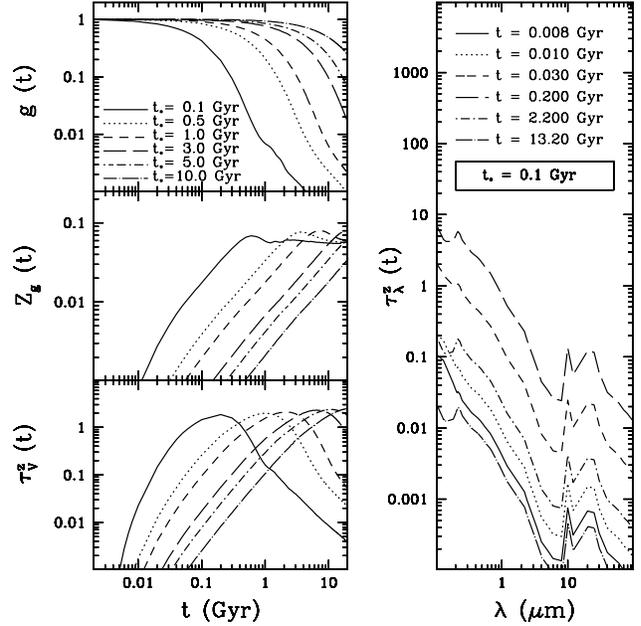}}
\hfill
\parbox[b]{87.5mm} {\caption{The left--hand side panel shows time 
evolution of the gas 
fraction, metallicity and optical depth for the closed box 
model respectively described by equations \ref{eqgas}, 
\ref{eqmet} and \ref{eqtau}. Coding for the different curves is 
given in the top panel. $t_*$ stands for the 
characteristic time scale for star formation as defined in 
equation \ref{eqsfr}. The right--hand side panel represents 
snapshots of
mean face-on (z-axis) optical depth as a function of wavelength 
for a galaxy with a star formation 
time scale of 0.1 Gyr according
to formula \ref{eqtau}. The different lines represent the
different snapshots at the time mentioned in the panel.
All the curves of both panels are computed with $f_{\rm H} = 1$.}
  \label{gas}}

\end{figure}

The results of such a chemical evolution model for star formation
histories with various $t_*$ are shown in the upper left panels of Fig.~\ref{gas}, where 
we plot the time evolution of the gas fraction $ g(t) $ defined as 
$g(t) \equiv M_g (t)/M_{tot} $ and the gas metallicity $Z(t)$.

\section{Computing Dust Spectra}

\subsection{Absorption}

To estimate the stellar flux absorbed by the ISM in a 
galaxy, one first needs to compute its optical depth. 
As in Guiderdoni and Rocca--Volmerange \cite*{GRV87}, and Franceschini et al.
\cite*{FZTMD91}, \cite*{FMZD94},  
we will assume that the mean face-on (z-axis) optical depth of the 
average gaseous disk at wavelength $\lambda$ and time $t$ is:

\begin{equation}
\tau_\lambda^\mathrm{z} (t)  = \left( {A_\lambda \over
A_\mathrm{V}} \right)_{Z_\odot} \left( {Z_g(t) \over Z_\odot} \right)^s 
\left({\langle N_{\sc\rm H}(t) \rangle \over
2.1~10^{21} {\rm~at~cm^{-2}}} \right) \, ,
\label{eqtau}
\end{equation} 
where the mean H column density (accounting for the presence of
helium) is written:
\begin{eqnarray}
\langle N_\mathrm{H}(t) \rangle \simeq 
6.8~10^{21} g(t) f_\mathrm{H} \, \, \mathrm{atoms \, cm^{-2}} \, \, .
 \nonumber
\end{eqnarray}

As noted by Guiderdoni \cite*{G87}, a galaxy with 
$g \simeq 20 \, \%$ has $\langle N_\mathrm{H} \rangle \simeq 
1.4~10^{21}$ atoms cm$^{-2}$,which is in fair agreement 
with the observational value for late--type disks,
so that $f_\mathrm{H} \simeq 1$ for normal spirals (corresponding
to $r_g/r_{25} \simeq 1.6$, see Guiderdoni and Rocca--Volmerange 
\cite*{GRV87} for details).
This is the value we will adopt in the ``standard'' case, 
but we will allow ourselves the possibility to increase the H 
column density to model ULIRGs for instance. In order 
to do this, we 
will just multiply the column density by a ``concentration'' factor 
$f_\mathrm{H}$ which is inversely proportional to the
surface occupied by the gas and accounts for the fact that 
star formation is more concentrated in starbursts, than
in normal spiral disks. Of course this is not completely satisfactory,
and $\langle N_\mathrm{H} \rangle$ should be 
physically linked to the size and mass of the gaseous
disk in order to get rid of $f_\mathrm{H}$. 
The way we estimate sizes and masses of gas disks will be detailed in 
a companion paper \cite{DG99}.

In equation \ref{eqgas}, the extinction curve depends on the gas 
metallicity $Z_g(t)$ according to power--law interpolations based on 
the Solar Neighbourhood and the Large and Small Magellanic Clouds, with 
$s=1.35$ for $\lambda < 2000$ \AA \, and $s=1.6$ for
$\lambda > 2000$ \AA \, (see  Guiderdoni and 
Rocca--Volmerange \cite*{GRV87} for details).  The extinction curve 
for solar metallicity $(A_\lambda /A_\mathrm{V})_{Z_\odot}$ is taken
 from Mathis et al. \cite*{MMP83} 
\footnote {For wavelengths shorter than 912 \AA \, a second term should be 
added to account for hydrogen absorption.
This term is of the type \cite{DSGN98}:
\begin{equation}
\langle N_{\sc\rm H}(t) \rangle \,  \sigma_{\scriptscriptstyle \rm H}
\left( {\lambda \over 912 {\rm ~\AA}} \right)^3 
\Theta (912 {\rm ~\AA} ) \, \nonumber ,
\end{equation}
$\Theta$ is the Heavyside function, and 
$\sigma_{\scriptscriptstyle \rm H} = 6.3 \times 10^{-18} 
\, \rm cm^{2}$ is the hydrogen
ionization cross section at the threshold.
This term represents the internal extinction of the H ionizing
continuum due to the presence of neutral hydrogen in galaxies
which has to be added to the external extinction of line--of--sight
photons due to intervening gas rich systems. As it is dependent of 
the clumpiness of the gas and 
represents a small fraction of the reprocessed photons  \cite{DSGN98},
we will not consider this extra absorption term in what follows.}. 

Examples of the behavior of the average face--on optical 
depth as computed using equation \ref{eqgas} can be read from 
Fig.~\ref{gas}. We note that with $f_{\rm H} = $1, the disks are
optically thin during most of their evolution.

\subsection{Geometry}

Once one has computed the average face--on optical thickness of a galaxy,
one needs to assume a given geometric distribution for the ISM, in
order to compute dust obscuration from dust extinction properties.

As in Dwek and V\'arosi \cite*{DV96}, we model a galaxy as an oblate ellipsoid
where absorbers (dust) and sources (stars) are homogeneously
mixed. Assuming a homogeneous absorption of the medium results in an 
overestimate
of the absorption of the UV photons and is discussed in more details
in Devriendt et al. \cite*{DSGN98}.  
Following the former authors we assume that the ``thickness-to-diameter'' ratio
of our disk--shaped galaxies is $\simeq 0.02$.

For spherical galaxies, Lucy et al. \cite*{LDGB89} generalized the analytic formula
giving obscuration as a function of optical depth
$\tau_\lambda^\mathrm{sph}(t)$ \cite{O89} 
to the case where scattering is
taken into account via the dust albedo $\omega_\lambda$.
Using Monte-Carlo simulations, Dwek and V\'arosi were able to show that 
the same formula is still valid provided one takes  
$\tau_\lambda(t) \simeq 0.193 \, \tau_\lambda^\mathrm{sph}(t) \simeq 2.619 \,
 \tau_\lambda^\mathrm{z}(t)$ as the effective optical thickness of
the disks at wavelength $\lambda$ and time $t$. The internal dust
obscuration, {\em i.e.} the {\em effective} extinction,
(averaged over inclination angle $i$) is then given by:
\begin{eqnarray}
\langle A_\lambda (t) \rangle_i &=& -2.5 \log_{10} \left[ {a_\lambda
\over 1 - \omega_\lambda + \omega_\lambda a_\lambda} \right] ,
\end{eqnarray}
where

\begin{eqnarray}
\nonumber
a_\lambda (t) &=& \left[\frac{3}{4 \tau_\lambda}
\left( 1- \frac{1}{2 \tau_\lambda^2}+(\frac{1}{\tau_\lambda}+\frac{1}{2
  \tau_\lambda^2}) \exp(-2 \tau_\lambda) \right) \right] .
\end{eqnarray}
\begin{figure}[htbp]

\resizebox{9cm}{!}{\includegraphics{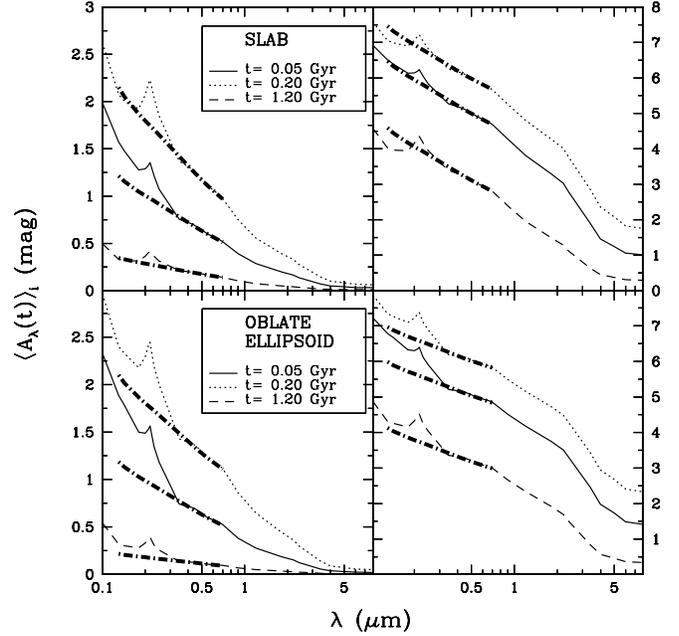}}
\hfill
\parbox[b]{87.5mm} {\caption{Time evolution of the 
obscuration for a galaxy with different geometries
for the distribution of dust and stars. The star formation 
time scale is 0.1 Gyr. For panels on the left hand side $f_{\rm H} = 1$,
whereas on the right hand side, $f_{\rm H} = 100$ (see 
text for a definition of $f_{\rm H}$). The thick 
dot-dashed curve is the polynomial fit given by Calzetti
et al. (1994), rescaled to our Q(5500).}
  \label{alam}}

\end{figure}

Figure \ref{alam} shows the 
comparison of $\langle A_\lambda (t) \rangle_i$
in our model with the commonly--used ``slab'' geometry.
In a slab geometry, stars and dust are just distributed
in the same infinite plane layer with the same vertical scale,
so that $\langle A_\lambda (t) \rangle_i$ is just given by:
\begin{eqnarray*}
\langle A_\lambda (t) \rangle_i &=& -2.5 \log_{10} \langle \left[
  {1 - \exp(-\sqrt{1-\omega_\lambda} \tau_\lambda^{\mathrm z} /\cos i) \over 
\sqrt{1-\omega_\lambda} \tau_\lambda^{\mathrm z} /\cos i} \right] \rangle_i
\end{eqnarray*}
where the albedo is phenomenologically taken into 
account through $ \sqrt{1-\omega_\lambda} $ \cite{GRV87}.
We also computed the absorption in a ``screen'' geometry, where the
dust layer is in front of the stars, and in the ``sandwich'' one, where the 
dust layer is sandwiched between two star layers.
We do not show results for these two alternative cases, because,
as shown by Franceschini \& Andreani \cite*{FA95}, and  Andreani \&
Franceschini \cite*{AF96} for a sample of LIGs and normal spirals, 
these geometries result in too much and not enough starlight
absorption respectively.    
On the other hand, the slab geometry (and the oblate ellipsoidal one
which we use as our standard in what follows) seems to yield
absorptions which are more in accordance with the data.   

As can be noticed in 
figure \ref{alam}, the results derived with the oblate ellipsoid model are quite 
similar to the ones obtained
from a slab model. For a concentration factor $f_\mathrm{H}=1$, {\em i.e.}
for the so called ``standard'' model which corresponds
to the average obscuration of normal spiral disks, the agreement
to the observational fit of Calzetti et al. \cite*{CKSB94} is fairly good in the 
optical/near--IR wavelength range (4000 to 8000 \AA). But both 
models overestimate the UV extinction by quite
a large factor ( about 1 magnitude at 1500 \AA), as can be seen on 
the left hand side of Fig.~\ref{alam}. This is due to the fact that 
the shape of the {\em obscuration} curve observed by Calzetti et al. \cite*{CKSB94} is
not a direct measurement of the extinction in a galaxy, but includes
geometrical effects. It just represents the 
relative obscurations (obscurations at other wavelengths are compared to 
the obscuration at 5500 \AA). For different geometries, these relative 
obscurations scale differently with the optical depth.
For instance, in the school case of the sandwich model only a
maximum value of half of the emitted light can be absorbed, corresponding
to a maximum value $\langle A_\lambda (t) \rangle_i = 0.75$. Therefore,
if absorption is already maximal at 5500 \AA, half
of the light emitted at 2000 \AA \, \, will also be absorbed, even though the optical depth 
is much larger at this wavelength! To sum things up, the global effect is very much 
geometry dependent, and if one picks the ``right'' geometry for 
the distribution of dust and stars, one can either reduce
the relative obscurations and smooth out any feature on the
absorption curve (``sandwich like'' effect), or increase the relative
obscurations and enhance the features (``screen like'' effect).

This effect is illustrated on the right--hand panels of 
Fig.~\ref{alam}, where the same
models of absorption are run, but with $f_\mathrm{H}=100$ which simply
means that the optical depth is multiplied by 100 
at all wavelengths (see eq.~\ref{eqgas}). 
One can see in Fig.~\ref{alam} that the ``slab'' geometry exhibits a clear proclivity to 
reduce the relative extinction between 5500 \AA \, and shorter
wavelengths, thus leading to a natural smearing of the extinction
curve features (carbon bump at 2000 \AA) and to a better 
agreement (on the entire wavelength range) with the fit given by 
Calzetti et al. \cite*{CKSB94} for a sample of obscured starbursts.
On the other hand, for the oblate ellipsoid geometry, the relative
absorption in the same range remains quite insensitive to the value
of the absolute optical depth. Considering our crude knowledge of 
the extinction curve of even close--by optically thin objects, and the
sensitivity of absorption to geometry illustrated above, we 
adopt the oblate ellipsoid geometry in what follows, keeping in mind 
that it might overestimate absorption at wavelengths shorter than
3000 \AA.

\subsection{Dust Model}

In the previous section we have shown how to compute the 
total stellar luminosity absorbed by the ISM. 
Next step is to re--distribute
this luminosity in the IR/submm window.

In order to be able to compute such an emission spectrum, 
what one needs is to model the dust
in the ISM. Our multi--component dust model is mainly 
based upon the model by D\'esert et al. \cite*{DBP90}. 
This model includes contributions from polycyclic aromatic
hydrocarbons (PAHs), very small grains (VSGs) and big 
grains (BGs).

Because of their small size ($\leq $ 1 nm), PAH molecules are out of
thermal equilibrium when excited by a UV/visible radiation field. 
Their temperature fluctuates and can
reach a value well above the equilibrium temperature, naturally
producing the bands at 3.3, 6.2, 7.7, 8.6 and 11.3 $\mu$m.
Both VSGs and BGs are made of carbon and silicates. The main
difference between these grains is a size difference, though
VSGs are probably dominated by carbon whereas BGs seem to
be dominated by silicates. VSGs have sizes between 1 and 10 nm. 
As a consequence, along with PAH
molecules, they never reach thermal equilibrium. Therefore, their
emission spectrum is much broader than a modified black body spectrum 
at a single equilibrium temperature. On the other hand, BGs, which
have sizes between 10 nm and 0.1 $\mu$m (almost) reach
thermal equilibrium and can be reasonably described by a modified black body
$\epsilon_\nu B_\nu(T_\mathrm{BG})$ with emissivity $\epsilon_\nu \propto \nu^m$ 
( where the index $m$ is such that $ 1 \leq m \leq 2$) and temperature
$T_{\rm BG}$. Hereafter we take $m = 2$.  

In contrast with Maffei \cite*{Mphd94}, we assume that
the population of BG is divided in two components:
\begin{itemize}
\item a cold component, with a fixed temperature $T_\mathrm{BG} = 17 $  K  
\item a ``starburst'' component which has the same
shape as the modified black body described above, but 
a higher $T_\mathrm{BG}$ to account for the fact that 
BGs receive extra heating from the radiation field in the 
star forming region. This component finds its justification
in observations of e.g. the typical local starburst galaxy M82
\cite{HGR94}, where the shape of the far--IR spectrum can be
explained by a higher temperature (up to 40 K instead of 17 K)
of BGs in thermal equilibrium with the radiation field.
\end{itemize}
Contrary to Maffei \cite*{Mphd94}, we do not take into account 
the possible fluorescence of PAH molecules at wavelengths shorter than 3 microns. 

\subsection{Assembling the Dust Puzzle}

The submm emission of galaxies, very sensitive to the spectral
characteristics of dust, is more poorly known than
the FIR. Submm fluxes of only a few tens of galaxies are available 
throughout the literature, and estimates of the amount of energy 
released in this range are strongly discrepant (e.g. Chini 
et al. \cite*{CKKM86}; Stark et al. \cite*{SDPHPLEC89}; 
Eales et al. \cite*{EWD89}; Chini and Kr\"ugel \cite*{CK93}). 
However, Stickel et al. \cite*{SB+al98} have shown that data from the
ISOPHOT shallow survey at 175 $\mu$m are inconsistent with a large
population of cold galaxies undetected by IRAS. 

Guided by all these data,
one can compute emission spectra of galaxies over the whole wavelength range
(near--IR to radio). We do this by adding the different
contributions from the populations of dust grains (see previous
subsection) to the corresponding post--absorption synthetic stellar spectra.
The dust dominated part of the spectra is picked among the spectral
library (see previous subsection) following a method
similar to the one developed by Maffei \cite*{Mphd94}. 

\begin{figure}[htbp]

\resizebox{9cm}{!}{\includegraphics{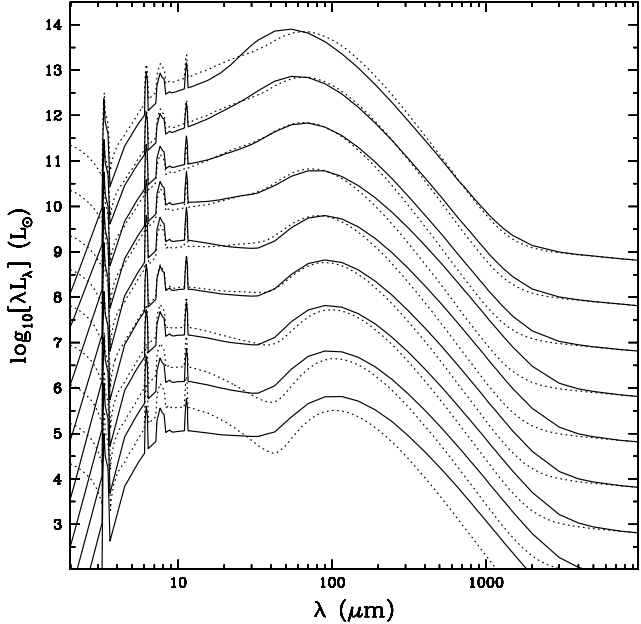}}
\hfill
\parbox[b]{87.5mm} {\caption{Spectral energy distribution 
in the IR window for galaxies with different total infrared
luminosities ranging from $10^6 L_{\odot}$ (bottom spectrum)
to $10^{14} L_{\odot}$ (top spectrum). Solid line is our model,
and the dotted curve is the model of Maffei (1994).}
\label{fir}}

\end{figure}

This method consists in using the observational correlations of the 
IRAS flux ratios 
12$\mu$m/60$\mu$m, 25$\mu$m/60$\mu$m and 100$\mu$m/60$\mu$m with
$L_\mathrm{IR}$ \cite{SN91} to determine the contribution
of each component of the dust model.  These correlations are
extended to low $L_\mathrm{IR}$ with the samples of
Smith et al. \cite*{SKHL87} and especially 
Rice et al. \cite*{RLS+al88}. We would like the reader to bear in mind
that these samples are rather small, so that the colour--luminosity
correlation is somewhat looser.
At submm wavelengths, the samples of Rigopoulou et al.
\cite*{RLRR96} and Andreani and Franceschini \cite*{AF96} are used 
to derive a correlation of the  350$\mu$m/60$\mu$m flux ratio with $L_\mathrm{IR}$.
As a result, we have now four color ratios which we use to
calibrate our four components. 

In Fig.~\ref{fir}, we show the SEDs
predicted by our model as a function of total
infrared luminosity. First of all, we notice an overall robustness
of the derived SEDs for $L_\mathrm{IR} \ge 10^8 L_\odot$, in the sense that
the shapes and emission maxima of our SEDs are very similar to those 
first derived by Maffei \cite*{Mphd94}. We have slightly more flux at 
wavelengths between 200 and 2000 $\mu$m for
galaxies with $L_\mathrm{IR} \le 10^{12} L_\odot$ which is the effect of
our different modelling of the BGs. As a matter of fact,
the cold part of the BG component is responsible for the excess
of flux (about a factor $\simeq 2$ at 850 $\mu$m) that is seen in 
Fig.~\ref{fir}.
For galaxies with $L_\mathrm{IR} \le 10^7 L_\odot$ we also have less
PAH emission and more mid--IR to submm emission due to the VSGs
and BGs. 

To further extend the wavelength range,
synchrotron radiation is added to the
spectra. This non--thermal emission is strongly correlated
with stellar activity and, as a consequence, with IR luminosity (see
e.g. Helou et al. \cite*{HSRR85}).  According to observations at 1.4 GHz, this
correlation is $L_\nu(1.4~GHz)=L_\mathrm{IR}/(3.75~10^{12} \times
10^q)$, where
$L_\mathrm{IR}$ is in W, $\nu_{80}=3.75~10^{12}$ Hz is the frequency at 80 $\mu$m and
$q\simeq 2.16$ is determined from observations. Then we assume that we can
extrapolate from 21 cm down to $\sim $ 1 mm with a single average slope 0.7,
so that $L_\nu=L_\nu(1.4~GHz)(\nu /1.4~GHz)^{-0.7}$.

This is the way we proceed for ``normal'' infrared galaxies as well
as for the models that are presented here unless the contrary 
is mentioned. For the
sample of ULIRGs of Rigopoulou et al. \cite*{RLRR96} we notice 
that the average slope is somewhat shallower, around 0.46 and the
parameter $q$ is more like $q\simeq 2.85$. As the sample is not 
large enough for this correlation to be firmly established, we
fit each ULIRG individually with specific $q$ parameter and slope,
in order to reproduce the observed radio flux. We also note 
that a steeper slope for some objects in the sample (e.g. Mrk 273)
probably indicates a non-negligible contribution in the radio flux 
coming from an active nucleus. Nevertheless, we assume that the 
dominant contribution in all the galaxies of this sample
comes from the star forming region, thus neglecting a possible pollution
by the active nucleus. This indeed seems to be a good
approximation, in view of results derived from a recent study of a sample
of 60 ULIRGs by Lutz et al. \cite*{LSRMG98}, as we expect 80 \% of the
objects to be powered by star formation in the IR.

\section{Building the Full Synthetic Spectra}

\begin{figure}[htbp]

\resizebox{9cm}{!}{\includegraphics{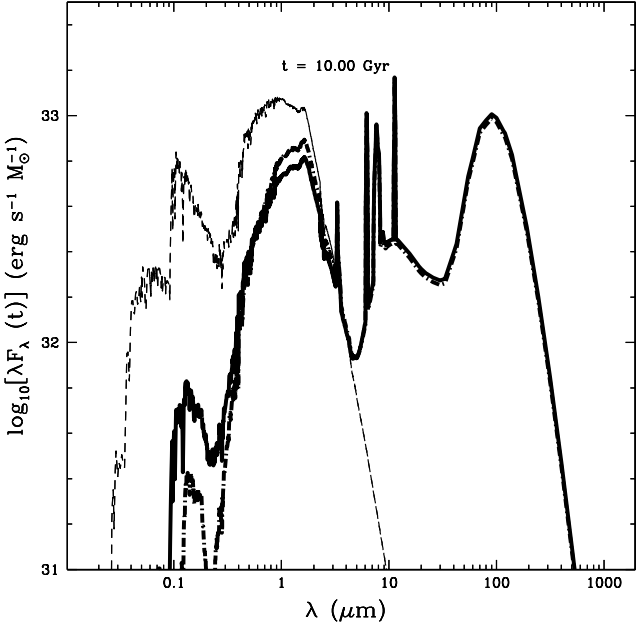}}
\hfill
\parbox[b]{87.5mm} {\caption{Snapshot of the full 
wavelength range synthetic
spectrum of a typical spiral galaxy with a star
formation time scale of 3 Gyr and $f_{\rm H} = 1$
taken at time $t = 10$ Gyr, corresponding
to $\tau_V^\mathrm{z} \simeq 1$ (see bottom left panel of 
Fig.~\ref{gas}). The initial gas mass 
available for star 
formation has been renormalized to $M_g(0)=M_{tot}=1~M_\odot$ 
(see eq \ref{eqgas}), the real initial value (which has a direct
influence on the IR spectrum) being $M_g(0)=10^{10}~M_\odot$. 
The different curves represent different 
geometries of dust and stars. Thick solid line is the homogeneous oblate 
ellipsoid mix, and thick dot--dashed line assumes a screen geometry. As a
guideline, we also plot (thin dashed curve) the spectrum without any
absorption.} 
  \label{comp_spec_1}}

\end{figure}

\begin{figure}[htbp]

\resizebox{9cm}{!}{\includegraphics{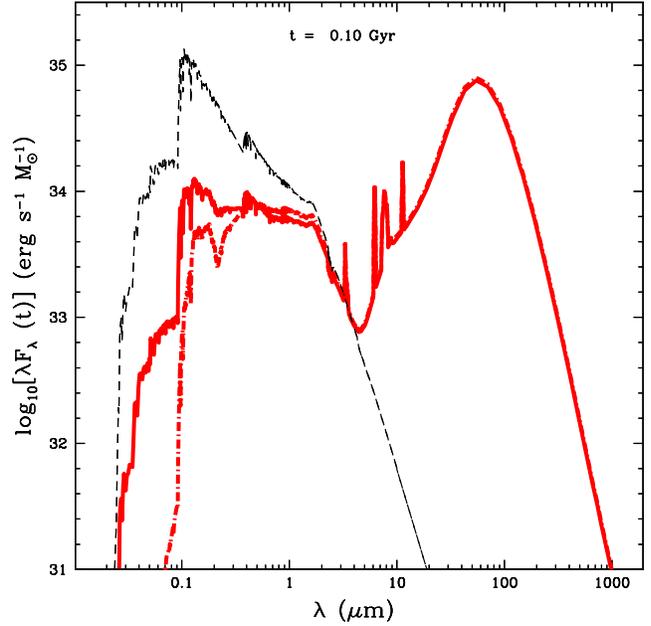}}
\hfill
\parbox[b]{87.5mm} {\caption{Snapshot of the full 
wavelength range synthetic
spectrum of a typical starburst galaxy with a star
formation time scale of 0.1 Gyr and $f_{\rm H} = 1$
taken at time $t = 10$ Gyr and corresponding
to $\tau_V^\mathrm{z} \simeq 1$ (see bottom left panel of 
Fig.~\ref{gas}). The initial gas 
mass available for star formation has been renormalized to
$M_g(0)=M_{tot}=1~M_\odot$ 
(see eq \ref{eqgas}), the real initial value 
being $M_g(0)=10^{12}~M_\odot$. 
The line coding is the same as in Fig.~\ref{comp_spec_1}.} 
  \label{comp_spec_2}}

\end{figure}

\begin{figure}[htbp]

\resizebox{9cm}{!}{\includegraphics{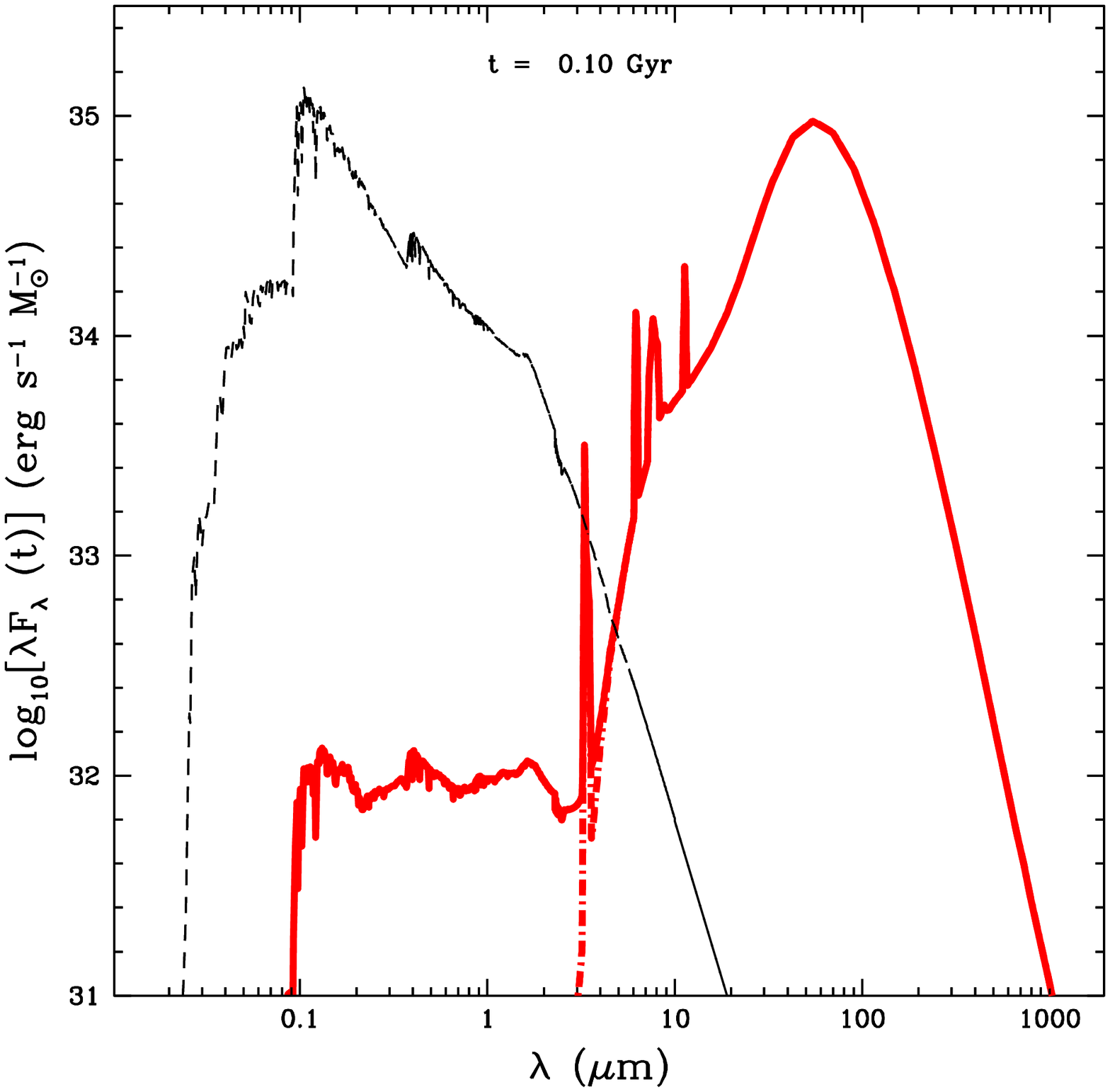}}
\hfill
\parbox[b]{87.5mm} {\caption{Same spectrum as in 
Fig.~\ref{comp_spec_2} but for $f_{\mathrm H} = 100$
thus corresponding to $\tau_V^\mathrm{z} \simeq 100$. 
The real initial gas mass available for star 
formation is still $M_g(0)=M_{tot}=10^{12}~M_\odot$ 
(see eq \ref{eqgas}). For the screen geometry (dot--dashed line), 
the level of the flux below 2 $\mu$m is negligible.} 
  \label{comp_spec_3}}

\end{figure}

Now that we have shown how to compute both the stellar
spectra and the dust spectra and how to link these
two components in a self--consistent way, we just piece 
them together in order to obtain the full galaxy spectra
from the far--UV to radio wavebands. This is shown in figures
\ref{comp_spec_1}, \ref{comp_spec_2} and \ref{comp_spec_3},
for different types of galaxies (normal spirals and 
starbursts with normal and high extinction) and for 
different kinds of geometries.In these figures, we also
overplot the original spectra as they come out 
from the spectral synthesis models without extinction.

The important thing to notice is that for normal spirals, the amount of 
luminosity absorbed by dust is
about half the luminosity produced by star formation, whereas for 
a starburst where the optical depth is calculated with the same
prescription, the luminosity released in the IR is already twice as large
as the luminosity emitted in the optical. This is mainly due to the fact 
that, in starbursts, star formation takes place on a time scale which is much 
shorter than in normal spirals, so that the stellar 
population has less time to grow old and it emits most of its flux
in the UV/optical window where absorption is higher. The other important
point to notice is that, when the objects remain marginally optically
thin (with $\tau_\lambda^{\mathrm z} \lesssim 1$), which is the case for the 
objects plotted in Fig.~\ref{comp_spec_1} and \ref{comp_spec_2}, the
UV/near--IR SEDs (except maybe at wavelengths $ < 2000$ \AA) are
fairly robust as they do not depend sensitively on the
geometry of the distribution of stars and dust.

For compact starbursts, the optical depth can increase 
by about two dex (we model these objects by putting our concentration
factor $f_\mathrm{H}=100$), and, as shown in Fig.~\ref{comp_spec_3}, two
important things happen. First of all, 
the ratio of the far--IR to optical fluxes is multiplied 
by a hundred, which was to be expected. Second, and less trivial
is that, depending on the geometry adopted, the shape of the optical
spectrum can undergo drastic changes. For instance, if one compares
figures \ref{comp_spec_2} and 
\ref{comp_spec_3}, it is obvious that, for an oblate ellipsoid
geometry, the shape of the stellar spectrum is quite independent of
the value of the concentration parameter $f_\mathrm{H}$, and the
optical SED just undergoes an attenuation of a couple of dex. 
On the contrary, for an extreme geometry like the screen model, 
the change is so drastic that the stellar spectrum that was present with
$f_\mathrm{H}=1$, has  totally disappeared for
$f_\mathrm{H}=100$. Of course, this is purely a test case, for which
the radiative transfer can be solved analytically, and the absorption
(neglecting inclination) varies exponentially with the optical depth, but nevertheless,
it is very illustrative of the sensitivity of the optical spectrum
to the dust distribution relative to stars. The explanation of this
dependency is that in the case of the
oblate ellipsoid geometry, we have a homogeneous mix of dust
and stars, so when the optical depth increases we still see the light
coming from the stars contained in the ``outer peel'' of the galaxy. 
On the contrary, for the screen geometry, the wall of dust in
front of the stars just thickens, and does not let any light go through at all. 

\begin{figure}[htbp]

\resizebox{9cm}{!}{\includegraphics{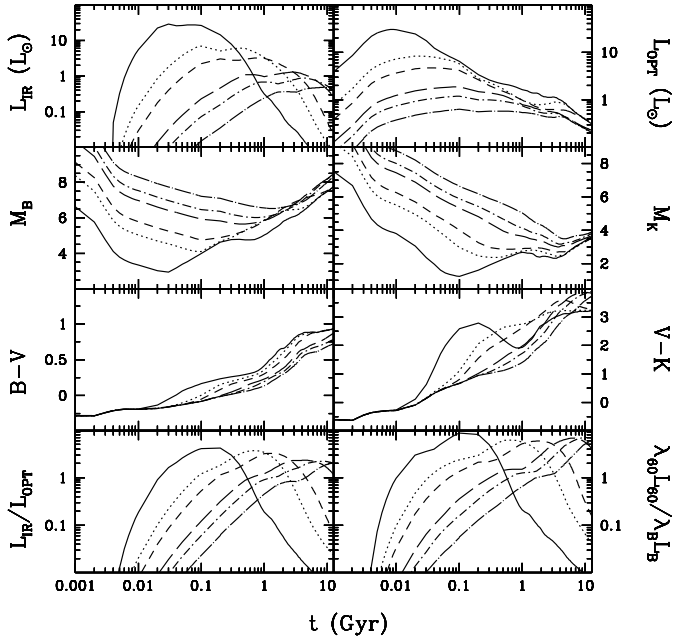}}
\hfill
\parbox[b]{87.5mm} {\caption{Time evolution of the 
luminosities and colors of a galaxy with different star 
formation time scales and $f_{\rm H} = 1$. 
Coding for the lines is the same as in 
Fig.~\ref{gas}, top left panel. The initial gas mass 
available for star formation is $M_g(0)=M_{tot}=10^{12}~M_\odot$ 
renormalized to $1~M_\odot$, and $L_\odot$ stands for solar bolometric
luminosity.}
  \label{lum_col_cl}}

\end{figure}

Another point we would
like to emphasize is the robustness of the 
dust emission spectrum. As a matter of fact, the 
level of flux for all wavelengths $> 10 \mu$m differs by less
than 20\% between ``compact'' starbursts (Fig.~\ref{comp_spec_2}) and
``normal'' starbursts (Fig.~\ref{comp_spec_3}) even if the absorption
of the starlight is two dex higher for a compact starburst. This is due to the fact
that the starlight is absorbed in a narrow wavelength range (typically
3 $\mu$m wide) and re--emitted over a much larger wavelength
range (typically 1000 $\mu$m wide). Moreover, the extra quantity 
of starlight absorbed by the compact starburst represents a very small
fraction of the bolometric luminosity $L_\mathrm{bol}$ of the object (and hence a very
small fraction of the total $L_\mathrm{IR} \simeq 2/3 L_\mathrm{bol}$).  
These combined effects have the dramatic consequence that one cannot predict the
visible spectrum of an ULIRG! 
However, one should be aware that the shape of the IR/submm SED is
very likely to be altered in compact starbursts, due to  
absorption of dust re--processed photons by other dust grains.
When the optical depth reaches values
as high as $\approx$ 100 in the V band, one can expect it to still be of
the order of 10 at 3 $\mu$m, and more or less of the same value up
to 30 $\mu$m because of the silicate absorption features around 10 and 25
$\mu$m (see Fig.~\ref{gas}). Our simple model does not account for such 
self--absorption explicitly, although one could argue that in principle it 
is more or less naturally accounted for, simply because of the way we 
build our spectral library from the phenomenological colors 
of IRAS galaxies. To sum things up, this effect should result in an important 
decrease of the strength of 
PAH features in highly obscured starbursts, which emit more than 90
$\%$ of their luminosity in the far--IR, and for this reason, it should 
be present in IRAS observations of ULIRGs which we use to build our spectral library.
We therefore believe that our conclusion about the overall robustness of the
IR SEDs still holds when dust self--absorption is explicitly taken into account.

\begin{figure}[htbp]

\resizebox{9cm}{!}{\includegraphics{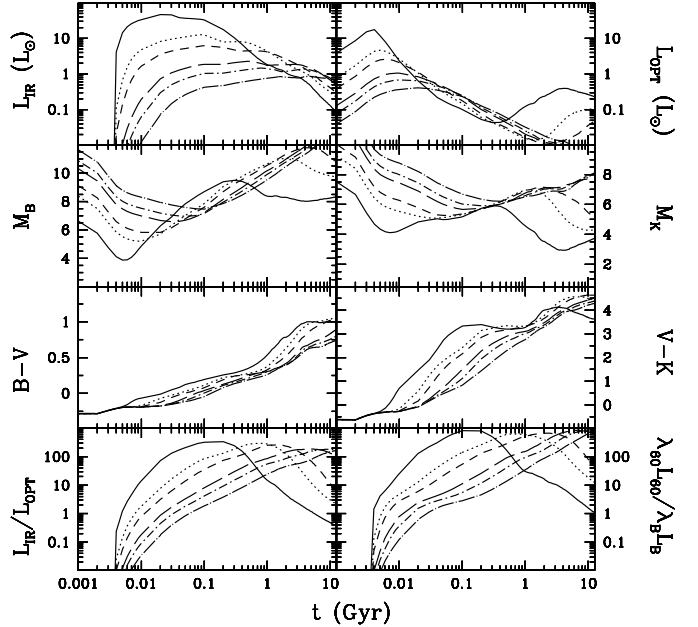}}
\hfill
\parbox[b]{87.5mm} {\caption{Same as in Fig.~\ref{lum_col_cl}
but with $f_{\rm H} = 100$.}
  \label{lum_col_di}}

\end{figure}

In Fig.~\ref{lum_col_cl} and \ref{lum_col_di}, we 
quantify these changes a bit further by plotting integrated
quantities and broad band colors. As expected, before 3
Myr, there is no IR flux, because no stars have yet exploded. 
So the ISM is not enriched with metals yet, and there is no 
dust. After 3 Myr, the metallicity starts growing, along with the dust
content and the IR luminosity of the galaxy, while in the same time 
more gas gets turned into stars. As the optical depth is a
function of both the metallicity and gas content, the IR luminosity 
reaches a maximum and then decreases. In fact, things are more complicated
because this behavior depends on the stellar emission spectrum. But, as shown
in the same plot (Fig.~\ref{lum_col_cl}), the optical luminosity has
a similar evolution, which is the result of a competition between 
stars forming at a rate which decreases with time because it is 
proportional to the gas content of the galaxy, and the ageing of
these same stars.

What one notices at first glance when one compares Fig.~
\ref{lum_col_cl} and \ref{lum_col_di}, is that the global behavior is
similar to what has previously been mentioned. The shapes of the spectra are
very much alike (mainly because the oblate ellipsoid geometry 
does not alter significantly the features and the extinction 
curve as previously discussed) in the sense that the colors
(especially B-V) are not noticeably different. Otherwise, the variations
of the plotted variables are those expected when one boosts
extinction: the IR luminosity increases, the optical luminosity drops,
and the absolute magnitudes in B and K increase.

The next step after checking the self--consistency of the model
consists in testing the spectra against real data in 
the full wavelength range. 

\begin{figure}[htbp]
\resizebox{9cm}{!}{\includegraphics{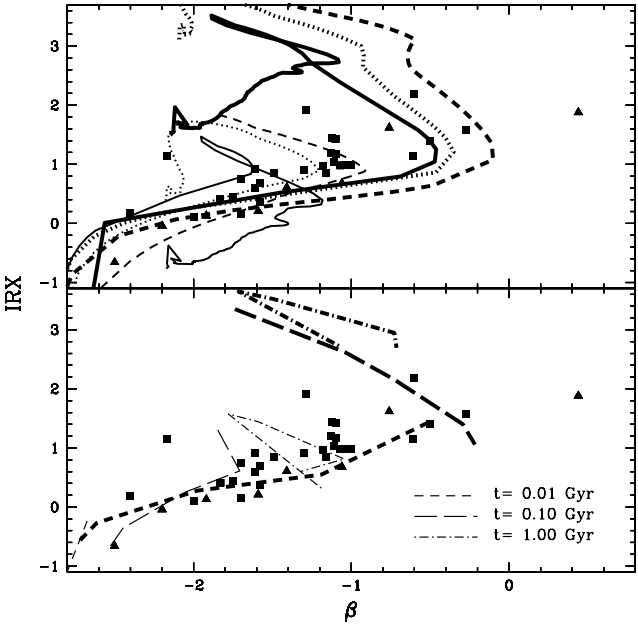}}
\hfill
\parbox[b]{87.5mm} {\caption{Flux
ratio IRX= $\log_{10}$ (FIR/F220W) as a function of UV spectral slope 
$\beta$. Data is from Meurer et al. (1997) and Kinney 
et al. (1994). Galaxies with different star 
formation time scales are 
overplotted.
Coding for the lines on top panel is : solid for 
$t_*= 0.1$ Gyr 
(thin line $f_{\rm H} = 1$
 and thick line $f_{\rm H} = 100$), dotted line for 
$t_*= 1.0$ Gyr, and dashes for $t_*= 10.0$ Gyr (the 
thickness coding remaining the same).
For the bottom panel, isochrones are plotted for 
different $t_*$ ranging from $0.1$ to $10$ Gyr. Coding for times
at which isochrones are taken is indicated in the panel.
Once again a thin line stands for $f_{\rm H} = 1$  
and a thick line for $f_{\rm H} = 100$. See text for 
more details and figure \ref{beta_IRX_evol} to 
estimate the starting values of $\beta$ and IRX in the bottom panel.}  
  \label{beta_vs_IRX}}
\end{figure}

\begin{figure}[htbp]
\resizebox{9cm}{!}{\includegraphics{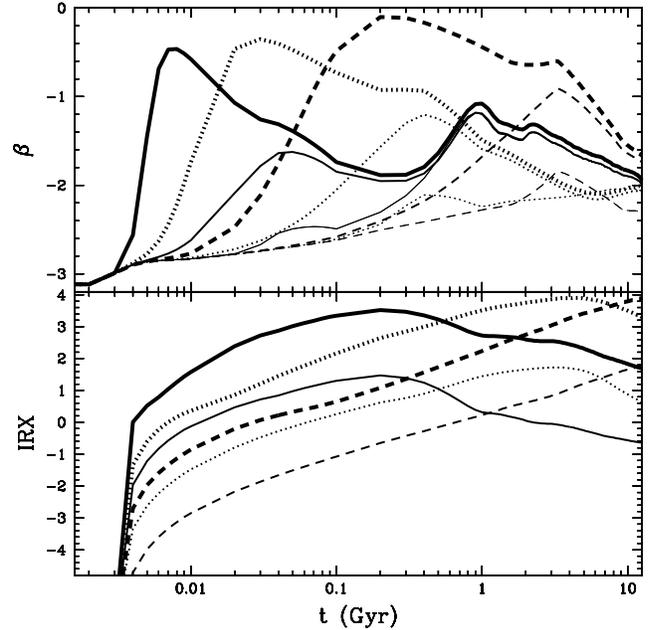}}
\hfill
\parbox[b]{87.5mm} {\caption{Time evolution of the flux
ratio IRX and UV spectral slope 
$\beta$ of galaxies with different star 
formation time scales. 
Coding for the lines on both panels is as follows: solid line
stands for  $t_*= 0.1$ Gyr, dotted line for 
$t_*= 1.0$ Gyr and dashes are for $t_*= 10.0$ Gyr.
In the top panel, the thin line represents models without extinction,
the thick line, models with $f_{\rm H} = 1$ and the very thick line, 
models with $f_{\rm H} = 100$. Obviously, for the bottom panel, there
is no thin curve, but the thickness coding remains the same in the
other two cases.}
  \label{beta_IRX_evol}}
\end{figure}

Fig.~\ref{beta_vs_IRX} and \ref{beta_IRX_evol} show how the models 
compare to 
observations in the UV and far--IR simultaneously.
The InfraRed eXcess (IRX) --- which is just the decimal logarithm
of the ratio of the far--IR flux (as estimated from IRAS observations)
and the flux measured in the F220W filter of the Hubble Space 
Telescope (see Meurer et al. \cite*{MHLKRG95}) ---, is shown both as a
function of time (Fig.~\ref{beta_IRX_evol}) and as a function of the 
UV spectral slope $\beta$ ---  which is defined as the power law
index of the UV continuum $F_\lambda (\lambda) \propto \lambda^\beta$ between 1250 \AA \, and 2600 \AA \, , 
(see Calzetti et al. \cite*{CKSB94} for details) --- (Fig.~\ref{beta_vs_IRX}).

There are a few important things to notice on these figures. 
In particular,
(upper panel of Fig.~\ref{beta_IRX_evol}), 
one can see that, for our geometry (oblate ellipsoid), the UV spectral slope remains 
negative whatever the value of $f_{\rm H}$, 
and that, in order to get (almost) flat spectra in
the UV ({\em i.e.} $\beta \simeq 0$) for starbursts, we need to invoke a 
higher extinction by setting $f_{\rm H}=100$.

It is striking that it is almost
impossible to obtain positive values of $\beta$ with reasonable 
values of the star formation rate and extinction. Of course, this 
statement depends on the geometry as well as on the IMF, but one would
have to come up with a very {\em ad--hoc} combination of these two 
parameters in order to change the sign of $\beta$. Furthermore,
as shown on the same figure (bottom panel), higher values  
of $\beta$ are not correlated with a strong IRX. They are reached 
{\em before} the IRX attains its maximum level.  This just means 
that if one demands that the geometry be the homogeneous 
oblate ellipsoid and the IMF be Salpeter, one would have to
advocate a column density at least a 1000 times larger in starbursts
than in normal spirals (the metallicity dependence becomes  
second order due to the fact that the peak in the time evolution
of $\beta$ appears very early) in order for the extinction to be 
sufficient to change the sign of $\beta$ ($f_{\rm H}=1000$ in
this case, and $\beta$ reaches a maximum value of 0.27).

However, as can be read in Fig.~\ref{beta_vs_IRX}, 
the vast majority of galaxies have negative
UV continuum slopes, and therefore, the models are able to span the whole data
range quite naturally.   
As a matter of fact, the spectral slope of the extinction--free starburst is 
$\beta \simeq -2.8 $ for $t < 0.02$ Gyr , in agreement with what is
generically used \cite{SAGDP99}. Furthermore, the predicted reddening is $ 0.04 <
E(B-V) < 0.4 $ (mainly depending on $f_{\rm H}$) for $t < $0.1 Gyr, in fair
agreement with what is derived in LBGs under the assumption of a universal
(averaged over 0.1 Gyr) $\beta \simeq -2.5 $ \cite{SAGDP99}.

\begin{figure*}[htbp]

\resizebox{18.5cm}{!}{\includegraphics{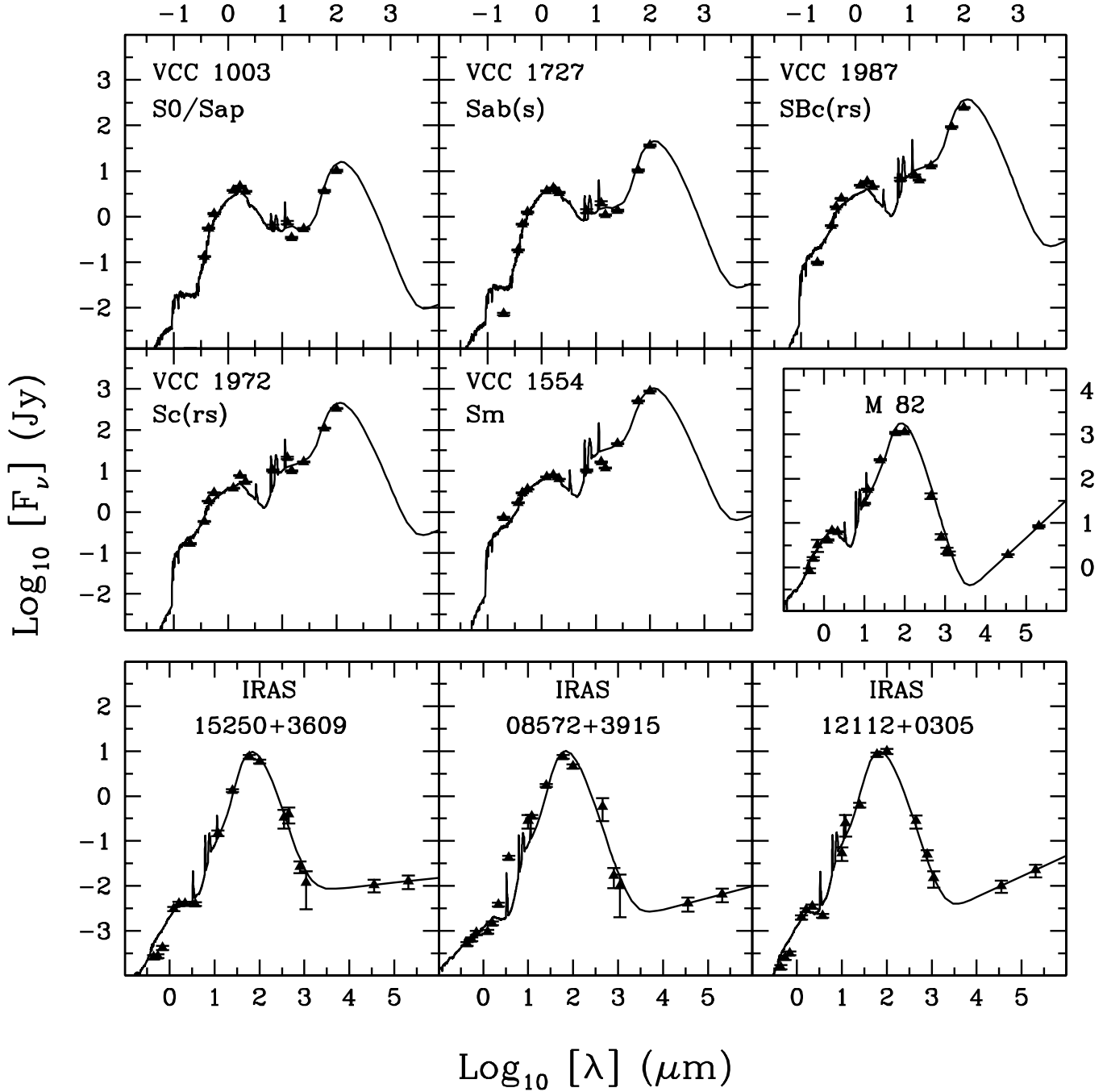}}
\hfill
\parbox[b]{18cm} {\caption{Best fit model (see text for 
details) for galaxies extracted from a sub-sample of the Virgo 
Cluster Catalogue (Boselli et al., 1998), and ULIRGs taken from the
sample described in Rigopoulou et al. (1996). As mentioned in the text,
this is illustrative of the ability of the model to
capture the characteristic features of the objects and there is
an important degeneracy between the different parameters 
($t_*$ and $t$) of the fit. Objects are ordered with increasing 
$ L_{\rm IR}$ from top to bottom and from left to right.}
  \label{fit}}
 
\end{figure*}

\subsection{Fitting Observed Galaxies} 

The next step is to test the ability of our models to 
describe multi--wavelength flux measurements of
a broad class of objects, from normal galaxies that emit 
approximatively $30$ \% of their bolometric luminosity in the IR,
to monsters which have $L_\mathrm{IR}/L_{\rm B} \simeq 100$.

To achieve this comparison, we simply select from the literature
a sample of galaxies for which fluxes in the UV, optical, IR
and submm are available. We also pick the sample in order to have 
galaxies with total IR luminosities (as estimated from IRAS 
observations, see e.g. Sanders and Mirabel \cite*{SM96}).
All the galaxies we show here are taken from samples described
in Rigopoulou et al. \cite*{RLRR96} for ULIRGs, and Boselli et al.
\cite*{B+al98} for normal spirals. We also add 
to the list the close--by LIRG M82 for which our multi--wavelength
data availability criterion is fulfilled (see Hughes et al. 
\cite*{HGR94} and references therein). 
We sum up the properties
of the selected objects in table~\ref{t_obj}. 

\begin{center}
\begin{table*}[htbp]
\caption {Properties of the different 
objects for which individual fits were achieved. Values given for the
absolute magnitudes are derived from the best fit
spectrum, {\em i.e.} the spectrum of the galaxy which age $t$ and
characteristic star formation timescale $t_*$ minimize the 
$\xi^2$. For $L_\mathrm{IR}$, values reported are obtained by integrating the spectra 
between 3 $\mu$m and 1000 $\mu$m, whereas, for $L_\mathrm{OPT}$, we
have integrated
the spectra between 100 \AA \, and 3 $\mu$m. SFR and $\tau_\mathrm{V}$ are 
indicative values that are derived from the best fit model.}
\begin{tabular}{|l||r|r|r|r|r|r|r|r|r|}
\hline 
Object Name  & $t_*$ (yr) & $t$ (yr) & D~(Mpc)  & $M_B$ & $M_K$
& $L_\mathrm{IR}$~($L_\odot$) & $L_\mathrm{OPT}$~($L_\odot$) 
& SFR~($M_\odot$ yr$^{-1}$) & $\tau_\mathrm{V}$ \\
\hline \hline
 & & & & & & & & & \\
VCC 0836  & 10$^{9}$ & 6.7 10$^{9}$ & 17 & -14.7 & -18.9 & 1.7 10$^8$ & 3.0 10$^8$ & 1.0 10$^{-2}$ & 0.6 \\
VCC 1003  & 10$^{9}$ & 1.4 10$^{10}$ & 17 & -14.2  & -18.2 &   2.1 10$^7$ & 1.8 10$^8$ & 2.3 10$^{-3}$ & 0.1  \\
VCC 1043  & 10$^{9}$ & 1.4 10$^{10}$ & 17 & -14.5  & -18.5 & 4.1 10$^7$ & 2.4 10$^8$ & 3.0 10$^{-3}$ & 0.1 \\
VCC 1554  & 10$^{10}$ & 3.2 10$^{9}$ & 17 & -16.0 & -18.8 & 6.9 10$^8$ & 5.2 10$^8$  & 1.0 10$^{-1}$ & 0.6 \\
VCC 1690  & 10$^{9}$ & 9.3 10$^{9}$ & 17 & -14.8  & -18.8 & 7.8 10$^7$ & 3.1 10$^8$ &  5.8 10$^{-3}$ & 0.3 \\
VCC 1727  & 10$^{9}$ & 1.1 10$^{10}$ & 17 & -14.5 & -18.6 & 5.0 10$^7$ & 2.5 10$^8$ & 3.8 10$^{-3}$ & 0.2 \\
VCC 1972  & 10$^{10}$ & 3.4 10$^{9}$ & 17 & -15.4 & -18.4 & 3.1 10$^8$ & 3.2 10$^8$ &  6.0 10$^{-2}$ & 0.6 \\
VCC 1987 & 10$^{10}$ & 3.3 10$^{9}$ & 17 & -15.3  & -18.2 & 2.7 10$^8$  & 2.9 10$^8$ & 5.4 10$^{-2}$ & 0.6 \\
M 82  & 5 10$^{8}$ & 9.0 10$^{7}$ & 4.2 & -19.3 & -22.9 & 6.0 10$^{10}$ & 1.6 10$^{10}$ & 9.1 10$^{0}$ & 2.2  \\
Arp 220  & 10$^{8}$ & 5.0 10$^{7}$ & 108 & -19.4 & -23.7 & 1.6 10$^{12}$ & 2.2 10$^{10}$  & 2.7 10$^{2}$  & 75 \\
Mrk 273  & 10$^{8}$ & 3.0 10$^{7}$ & 222 & -20.7 & -25.0 & 2.6 10$^{12}$ & 7.1  10$^{10}$ & 4.2 10$^{2}$ & 34 \\
UGC 05101  & 10$^{8}$ & 2.0 10$^{7}$ & 240 & -20.9  & -25.3 & 1.7 10$^{12}$  & 9.3 10$^{10}$ & 3.0 10$^{2}$ & 15 \\
IRAS 05189-2524 & 10$^{8}$ & 3.0 10$^{7}$ & 256 & -21.3 & -26.1 & 2.7 10$^{12}$  & 1.6 10$^{11}$ & 4.6 10$^{2}$ & 34 \\
IRAS 08572+3915 & 10$^{8}$ & 2.0 10$^{7}$ & 349 & -20.7  & -23.8  & 2.4 10$^{12}$ & 4.7 10$^{10}$ & 4.5 10$^{2}$ & 15 \\
IRAS 12112+0305 & 10$^{8}$ & 5.0 10$^{7}$ & 435 & -20.4  & -24.6 & 3.4 10$^{12}$ & 5.1 10$^{10}$ & 6.6 10$^{2}$ & 56 \\
IRAS 14348-1447 & 10$^{8}$ & 3.0 10$^{7}$ & 495 &  -20.9 & -25.0 & 4.0 10$^{12}$  & 7.8 10$^{10}$ & 6.5 10$^{2}$& 34 \\
IRAS 15250+3609 & 10$^{8}$ & 3.0 10$^{7}$ & 319 & -20.2 & -24.4 & 2.0 10$^{12}$ & 4.3 10$^{10}$  & 3.3 10$^{2}$& 34 \\
& & & & & & & & & \\
\hline
\end{tabular}
\label{t_obj}
\end{table*}
\end{center}

The parameters that we consider free when we perform the fit are 
the star formation characteristic time--scale $t_*$, and the age of 
the galaxy $t$.
The fits are obtained by a standard $\chi^2$ minimization procedure,
although we are aware that it has not much statistical significance, 
mainly because the 
number of data points used to perform the fit is too small 
(at most 17 fluxes at different wavelengths). 
Moreover, the assumption that errors are distributed normally
in the statistical sense (which is a requirement for a real $\chi^2$
procedure to be valid) probably does not hold.
Typical values of the parameters $t$ and $t_*$ for the objects 
represented in the different panels of Fig.~\ref{fit} can be read
in table~\ref{t_obj}, along with typical properties derived from the
best fit model. For the remaining objects of table~\ref{t_obj}, 
spectra are available from the authors upon request.

The results of the method we use to obtain the fit  
can be summarized in the following way, depending on the
nature of the fitted objects:
\begin{itemize}
\item In the case of the ULIRG sample (in which we include M82 which
  is not an ULIRG), we find that 
for most objects, an old population (with a typical age of 10 Gyrs) with 
the same extinction as the starburst population ($f_{\rm H}=100$) 
is required to account for the excess
of flux in the near--IR (around 1-2 $\mu$m). This is due to 
the fact that even in our simple modelling, we cannot assume that the 
measured flux is entirely due to the burst, and we have to take into account the 
previous star formation history of the galaxy. Clearly, in some cases,
there is a degeneracy between $t_*$ and this population. This can be
understood if one keeps in mind that increasing $t_*$ simply means
that the stellar population builds up more slowly, and consequently has 
time to grow old and therefore emits more flux in the near--IR. \\

\item For the sample of normal spirals, we allow the same parameters $t_*$
and $t$ to vary, but we take $f_{\rm H}=1$. We then add in a few cases
a heavily extinguished starburst ($f_{\rm H}=100$) to account for the 
missing far--IR flux. Once again, this is due to the fact that having
just one characteristic time scale for star formation in our model
implicitly assumes that, for these objects, all the measured flux comes
from the quiescent continuous star formation history of the galaxy,
which might describe the average properties of a sample of
galaxies, but is too crude when one deals with individual objects in
an IR--selected sample.

\end{itemize}

\begin{figure}[htbp]

\resizebox{9cm}{!}{\includegraphics{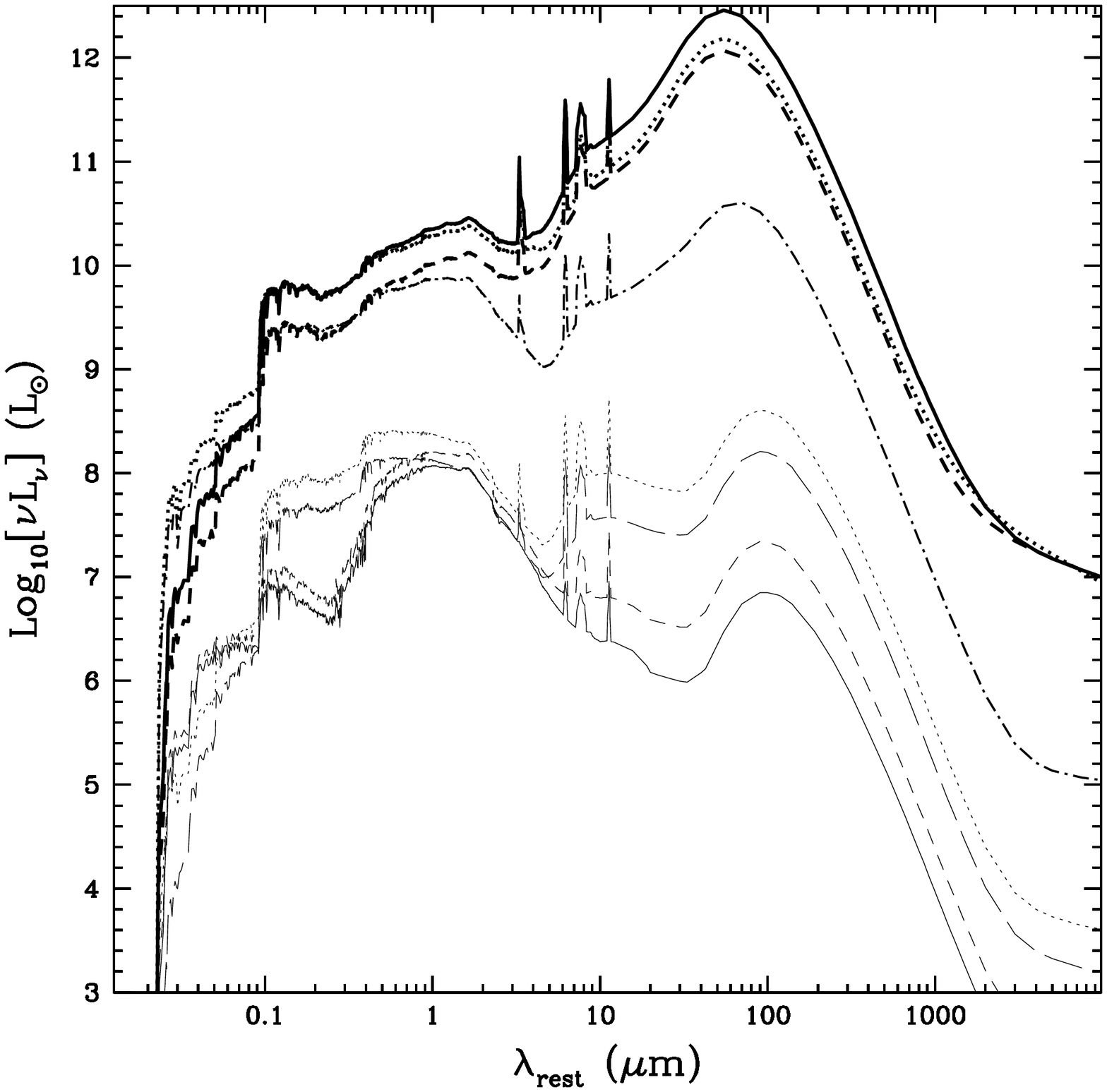}}
\hfill
\parbox[b]{87.5mm} {\caption{Sequence of galaxies 
taken from our best fit models (see table \ref{t_obj}) 
with different total IR luminosities
as in Sanders and Mirabel 1996. Distances to spiral galaxies
in the Virgo Cluster are taken to be 17 Mpc.
For the ULIRGs, distances are estimated by 
Sanders et al. (1988), rescaled to $H_o =$ 50 km s$^{-1}$ Mpc$^{-1}$.
Coding for the lines gives the name of each object: thin solid
line: VCC 1003; thin short--dashed line: VCC 1727; thin dotted line:
VCC 1554; thin long--dashed line: VCC 1987; thick dot--dashed line: M82;
thick short--dashed line: Arp 220; thick dotted line: IRAS 15250+3609;
and thick solid line: IRAS 12112+0305.}
  \label{san_mir}}
 
\end{figure}

From all the individual objects fitted, we were then able to extract 
a sequence of galaxies with different total IR luminosities. This is
shown in Fig.~\ref{san_mir} (in the spirit of Sanders \& Mirabel \cite*{SM96}) . 

One can see in this figure that,
as the IR luminosity increases, the near--IR to UV spectral slope
becomes shallower, and the characteristic features like the 
4000 \AA \, break are smeared out. This is the result of a combination
of lower absorption and older stellar population for the sample
of normal spirals as opposed to higher extinction and younger stellar
population for the ULIRGs. However, the figure is somewhat misleading,
because one could also see a luminosity sequence  
in the UV/near--IR SEDs that correlates with $L_\mathrm{IR}$.
As pointed out earlier in the text, there is  no such sequence,
for the simple reason that the UV/near--IR SED is {\em extremely} 
sensitive to the relative distribution of stars and dust in heavily
obscured objects. For instance, Arp 220 has
a SED in this wavelength range similar to M82 and its $L_\mathrm{IR}$
is about 100 times larger (see table \ref{t_obj}). 
Finally, the peak of the infrared emission
gets shifted towards higher temperatures or equivalently towards shorter
wavelengths as the total IR luminosity of the object increases.

\section{A Guide to High--z Galaxies}

The model described in the previous sections is designed
to match available multi--wavelength data of local galaxies.
In this section, we use the template spectra derived
from the model to make predictions
for objects which look like these local galaxies.

\begin{figure}[htbp]

\resizebox{9cm}{!}{\includegraphics{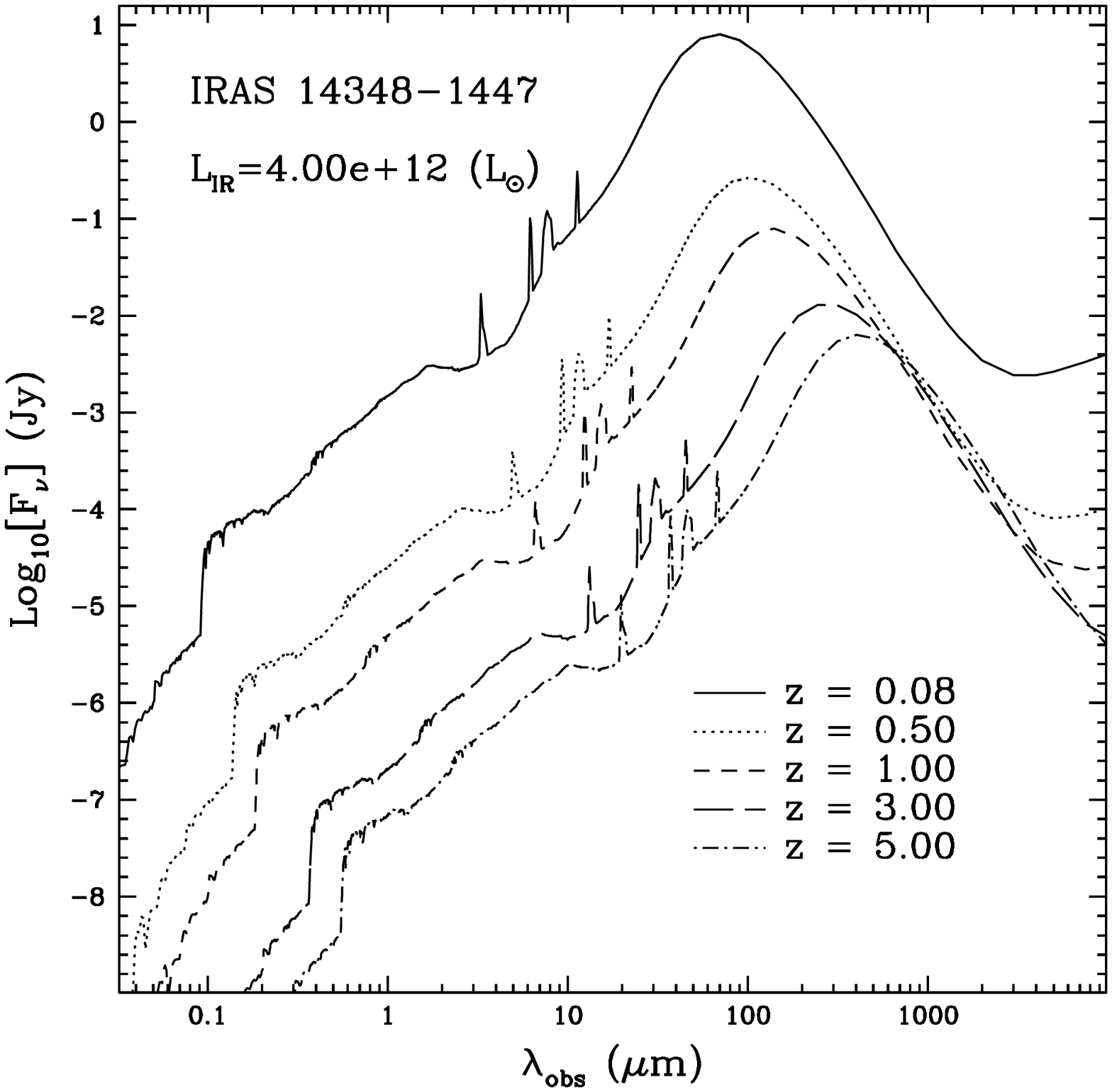}}
\hfill
\parbox[b]{87.5mm} {\caption{Observer--
frame spectrum of best fit model of the ULIRG galaxy
IRAS 14348-1447 at increasing redshifts, for a cosmology
where $H_o =$ 50 km s$^{-1}$ Mpc$^{-1}$, $\Omega_o = 1$ and $\Omega_\Lambda = 0$. }
  \label{ulred}}

\end{figure}

More specifically, we show in Fig.~\ref{ulred} how the brightest
ULIRG from our sample would look at different redshifts.
We would like to draw the attention of the reader to the crucial
effect of the so--called negative k--correction at submm wavelengths which
makes such objects as luminous at $ z \simeq 5 $ as at
$ z \simeq 0.5 $ for wavelengths ranging between 400 and 900 $\mu$m. 
\begin{figure}[htbp]

\resizebox{9cm}{!}{\includegraphics{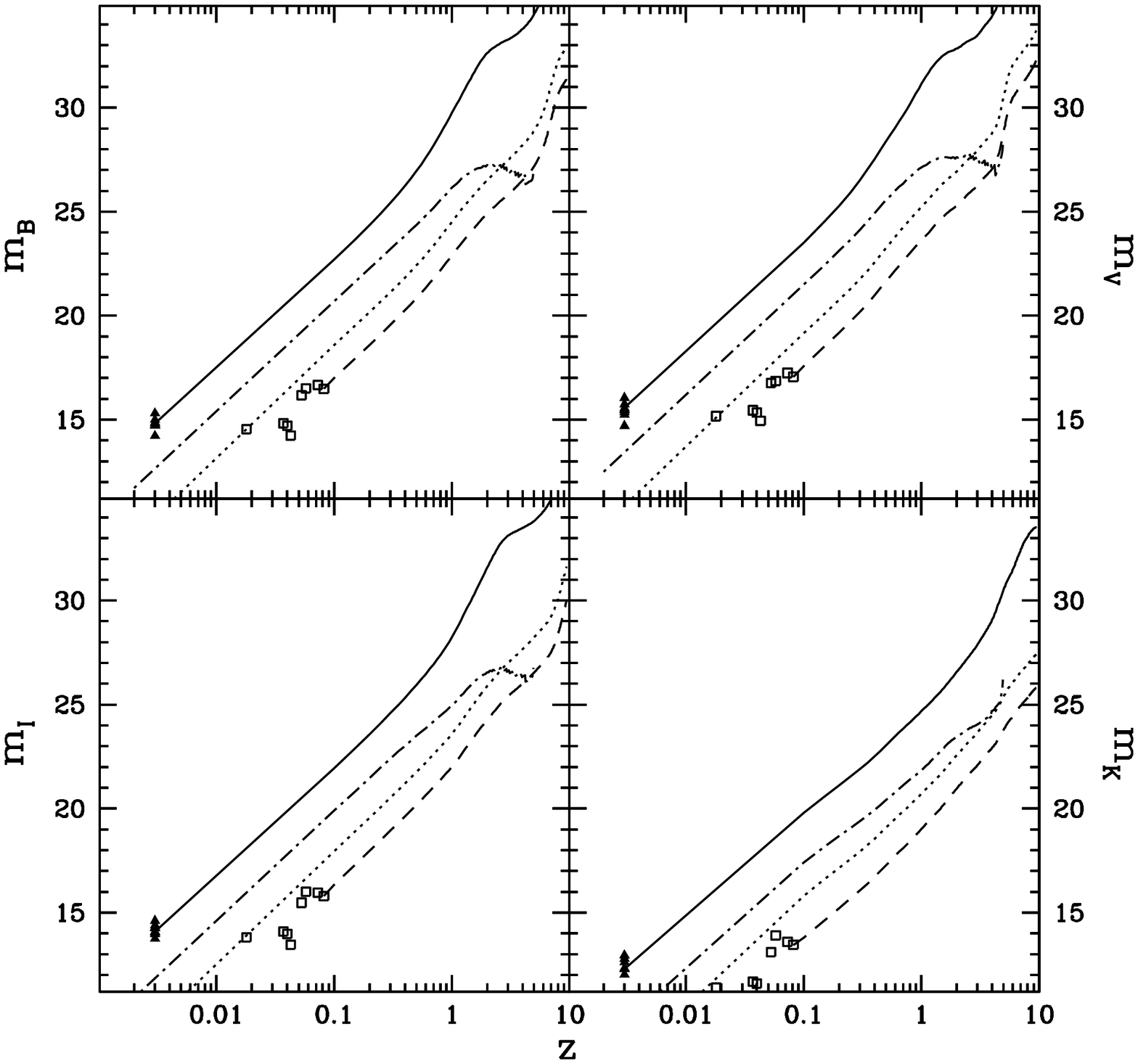}}
\hfill
\parbox[b]{87.5mm} {\caption{Apparent magnitude for 3 objects,
 the ``normal'' spiral VCC 836 of the Virgo Cluster (solid line), the
 close--by LIRG M82 (dotted line),
 and the ULIRG IRAS 14348-1447 (dashed line), as a function of
 redshift, for an Einstein--de Sitter cosmology
where $H_o =$ 50 km s$^{-1}$ Mpc$^{-1}$, $\Omega_o = 1$ and
$\Omega_\Lambda = 0$. Symbols are galaxies from table \ref{t_obj}. Dot--dashed line is a
model spiral with $t_* = 3$ Gyr and an initial gas mass available
for star formation of $M_g = 10^{10}~M\odot$ that formed at redshift
5, and for which the evolution correction is included. }
  \label{kcopt_std}}

\end{figure}

\begin{figure}[htbp]

\resizebox{9cm}{!}{\includegraphics{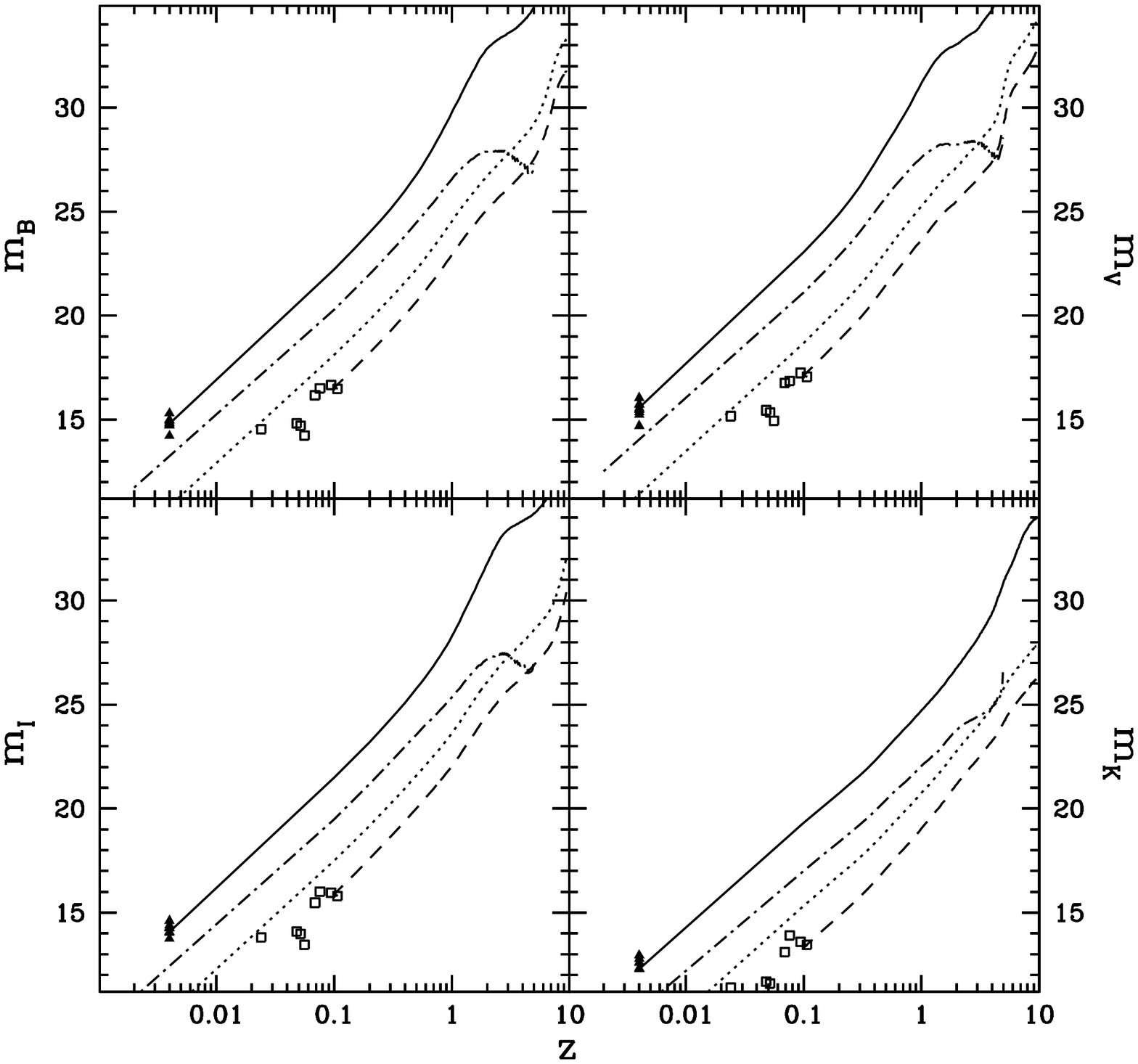}}
\hfill
\parbox[b]{87.5mm} {\caption{Same as Fig.~\ref{kcopt_std} 
for a flat cosmology
where $H_o =$ 65 km s$^{-1}$ Mpc$^{-1}$, $\Omega_o = 0.3$ and
$\Omega_\Lambda = 0.7$.}
  \label{kcopt_lam}}

\end{figure}

In Fig.~\ref{kcopt_std} and \ref{kcopt_lam}, we show 
optical/near--IR properties that
various classes of local objects (from small normal spirals to
bright ULIRGs) would have, if they were located at high redshift.
The two different pictures represent two different cosmologies
(Standard CDM and $\Lambda$CDM, see figure captions for details), 
and the differences induced by this are of the
order of 0.5 magnitudes in the various filters. Also the
dot--dashed line shows the evolution effects one expects to measure
on the spectrum of a spiral galaxy which forms at redshift 5 with an
initial gas mass $M_g(0) = 10 ^{10}\, M_\odot $.

\begin{figure}[htbp]

\resizebox{9cm}{!}{\includegraphics{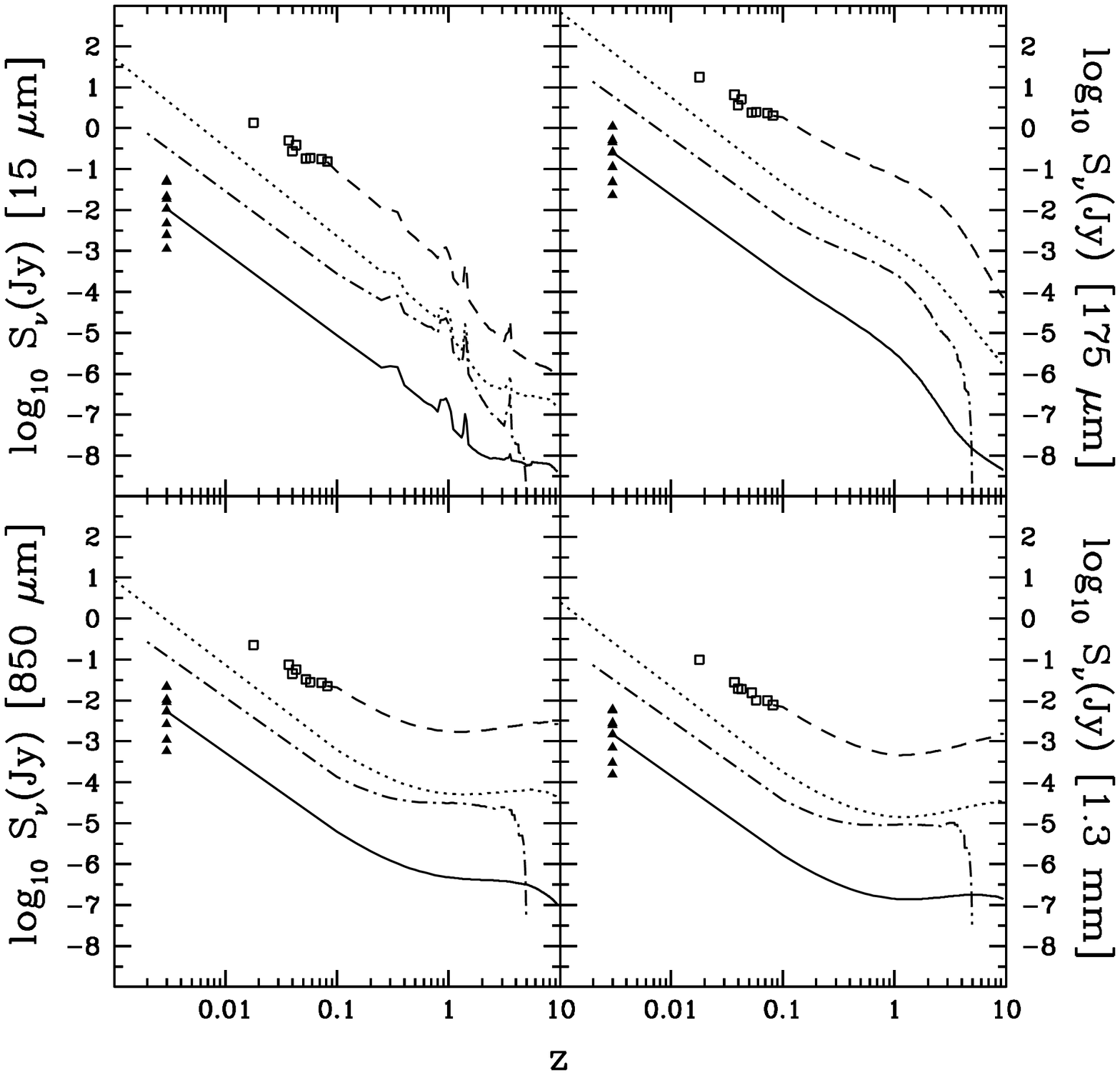}}
\hfill
\parbox[b]{87.5mm} {\caption{Apparent flux for the same objects in 
different IR/submm wavebands, as a function of redshift for the
Einstein--de Sitter
cosmology. Coding for the lines is the same as in Fig.~\ref{kcopt_std}}
  \label{kcinf_std}.}

\end{figure}

\begin{figure}[htbp]

\resizebox{9cm}{!}{\includegraphics{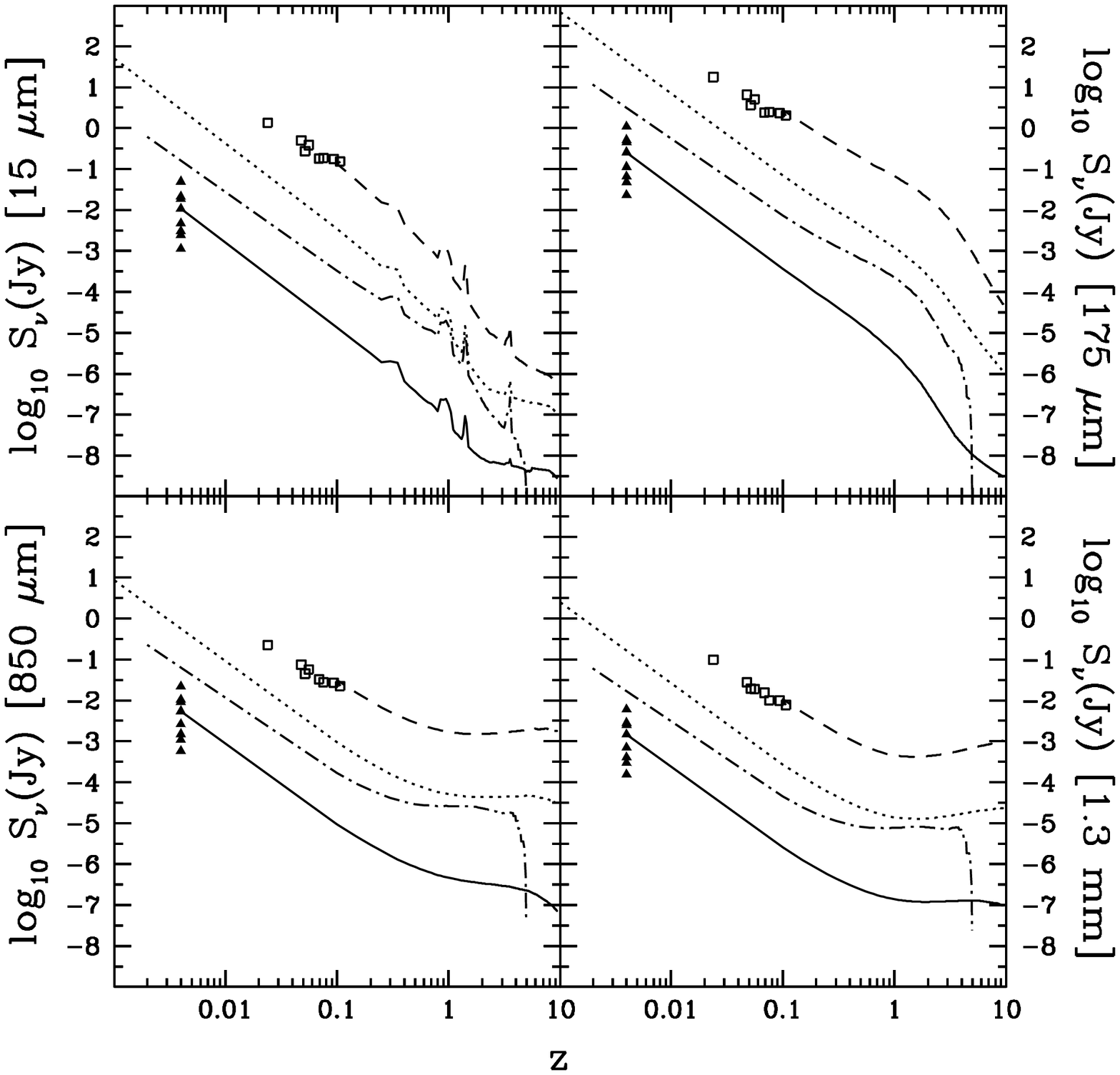}}
\hfill
\parbox[b]{87.5mm} {\caption{Same than Fig.~\ref{kcinf_std}
for the flat cosmology with a non--zero cosmological constant. Coding for the lines
is the same as in Fig.~\ref{kcopt_std}.}
  \label{kcinf_lam}}

\end{figure}

Fig.~\ref{kcinf_std} and
\ref{kcinf_lam} are far--IR/submm properties of the same classes 
of objects. Here again, the important thing to notice is that 
the change in fluxes due to the different cosmologies is quite
small (less than 30 \%) at any wavelength.

\begin{figure}[htbp]

\resizebox{9cm}{!}{\includegraphics{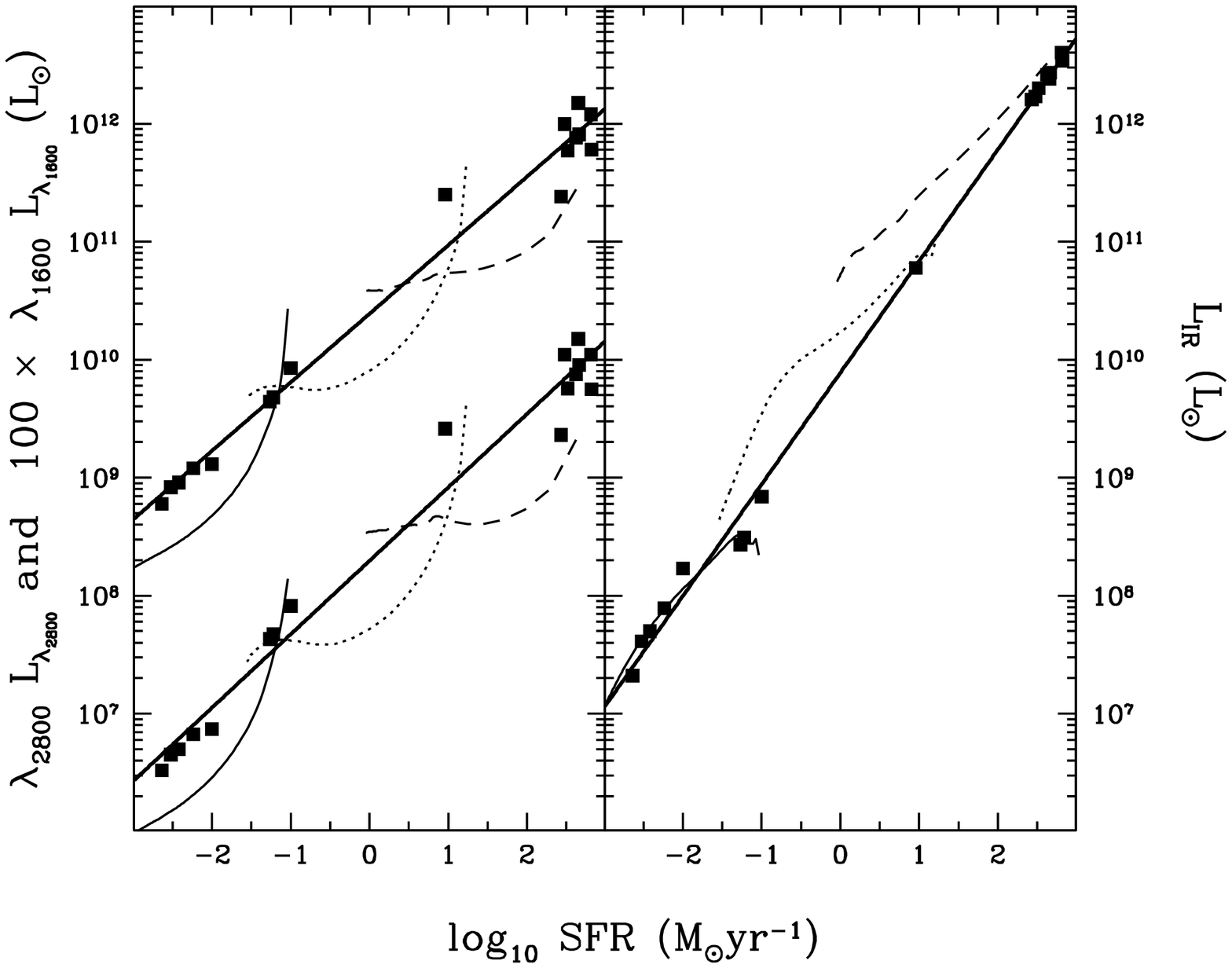}}
\hfill
\parbox[b]{87.5mm} {\caption{Star formation rate 
against luminosity at 2800 \AA \, and 1600 \AA \, (left panel), and total infrared 
luminosity (right panel). Symbols represent quantities derived from
our best fit models, some of which can be read from
table~\ref{t_obj}. For clarity, both these quantities and curves at 1600 \AA \,
have been arbitrarily shifted up by two decades. 
Coding for the lines is the following: solid for 
a galaxy with initial gas mass $M_g(0)= 10^8 \, M_\odot$,
$t_*= 1 $ Gyr and $f_{\mathrm H} = 1$; dotted line for a galaxy 
with initial gas mass $M_g(0)= 10^{10} \, M_\odot$,
$t_*= 0.5 $ Gyr and $f_{\mathrm H} = 10$; dashed line for a galaxy 
with initial gas mass $M_g(0)= 10^{11} \, M_\odot$,
$t_*= 0.1 $ Gyr and $f_{\mathrm H} = 100$. The curves show 
time evolution for the above mentioned objects
for ages ranging from $t=$ 0.01 (maximum star formation rate and
luminosities) to 15 Gyr (minimum star formation rate and
luminosities).
The straight lines represent least square fits with values for
parameters given in the text.}
  \label{sfr_l}}

\end{figure}

Eventually, Fig.~\ref{sfr_l} shows the time evolution of UV luminosities
$\lambda_{2800} L_{\lambda_{2800}} $,
$\lambda_{1600} L_{\lambda_{1600}} $ (left panel) and IR luminosity 
$ L_{\rm IR} $ (right panel), considered as functions of the star formation
rate. We remind the reader that the SFR time evolution 
is driven by equation~\ref{eqsfr}, and that our galaxies are
considered as isolated objects (``closed box'' approximation).
The interesting thing to note is the tightness of the correlation
between $ L_{\rm IR} $ and SFR over the five orders of magnitudes 
spanned by the galaxies we picked. 

One could point out that there seems to be 
a correlation between the UV luminosities and SFR too,
but, looking carefully at Fig.~\ref{sfr_l}, this holds only for 
SFR smaller than a few $ M_\odot$ yr$^{-1}$. This is obviously
due to the high extinction of all the objects that are massively
forming stars (SFR $ > $ 100 $M_\odot$ yr$^{-1}$): their 
UV luminosities are not significantly different
from objects with more normal SFRs (up to 10 $M_\odot$ yr$^{-1}$).
However, we emphasize that this should be viewed 
as a theoretical correlation, because the points plotted on
Fig.~\ref{sfr_l} are deduced from the best fit model.
Nevertheless, the mathematical expression
we get from simply fitting this ``data'' with a power law is 
$ L_{\rm IR}~(L_\odot)~=~7.7~\times~10^{9}~({\rm SFR}~/~M_\odot~{\rm yr}^{-1})^{0.95}$
where the exponent of the SFR is remarkably close to 1.
The same fitting method for the UV luminosities
yields :
$\lambda_{2800}~L_{\lambda_{2800}}~(L_\odot)~=~2.0~\times~10^{8} 
~({\rm SFR}~/~M_\odot~{\rm yr}^{-1})^{0.62}$  
and
$\lambda_{1600}~L_{\lambda_{1600}}~(L_\odot)~=~2.4~\times~10^{8}~ 
({\rm SFR}~/~M_\odot~{\rm yr}^{-1})^{0.58}$
with exponents fairly different from unity. Furthermore the scatter
is definitely more important.  
All this points out that it is indeed very important to have a fair estimate 
of the $L_{\rm IR}$ of high--redshift galaxies if one wants to
derive realistic values for their SFRs.

\section{Conclusions}

In this paper, we have described a simple model which enables us to reproduce the far--UV to 
submm SEDs of observed galaxies, with a possible extension 
to radio wavelengths. The specificity of our model is 
the fact that it self--consistently links starlight and light reprocessed
by dust to derive the global spectra of
galaxies. We hereafter summarize the main assumptions
of our work that attempts to be as phenomenological as possible.

To obtain SEDs of synthetic stellar populations, we couple spectrophotometric 
and chemical evolutions in a simple way, under the assumptions of a constant 
IMF (we take Salpeter IMF) and no gas inflows/outflows (closed--box). 
Once stellar SEDs are computed, we use a phenomenological extinction curve
that reproduces the observed metal--dependent trend in the SMC ($Z =
1/5 \, Z_\odot $), LMC ($Z =1/3 \, Z_\odot$) and Milky Way ($Z = \, Z_\odot$). The optical thickness can be 
computed e.g. under the assumption of no gas inflows/outflows. A simple oblate
geometry where stars and dust are homogeneously mixed finally gives the
obscuration curve. The absorbed luminosity is redistributed at IR/submm 
wavelengths with a phenomenological model involving various dust components 
that reproduce the local correlation of IRAS and submm flux ratios with 
IR luminosity. The stellar and dust SEDs are then coupled. In this simple 
approach, the predictions depend on the SFR timescale, galaxy age, and size
of gas disk that is simply parameterized with the fudge factor
$f_{\mathrm H}$.   

We have shown that the results derived separately for the UV/optical part
of the spectrum and the far--IR/submm luminosity distribution are 
robust in the sense that they are very close to the results obtained
independently by other authors. The predicted stellar SEDs are in good overall 
agreement with other predictions available in the literature, in spite of 
different origins for the stellar data. The differences are mainly due to 
the fact that the late stages of stellar evolution (TP--AGB and post--AGB) 
are not included in our modelling. The dust SEDs are in good agreement with 
those of Maffei \cite*{Mphd94}, who does not use submm data and takes slightly 
different components for the dust. The differences amount to at most a factor 2,
and decrease with increasing luminosity.

Furthermore, we have demonstrated that the results of the observational and 
theoretical works on extinction by Calzetti et al. \cite*{CKSB94} and Meurer et al. 
\cite*{MHLKRG95} can be naturally interpreted in the 
framework of our model as geometrical effects in the relative dust/star 
distributions. Indeed, in the optically--thick regime with 
optical depth  $\tau_\mathrm{V}^{z} \simeq 100$, we observe a flattening
and smearing of the $\langle A_\lambda \rangle_i$ curve --- as reported 
by Calzetti et al. \cite*{CKSB94} --- that depends on the way 
dust and stars are distributed.  
The obscuration curve is in good agreement with Calzetti's, though it is 
less ``grey''. A possible explanation is that Calzetti's starbursts are 
more metal--rich on average, than our model starbursts that start from 
zero metallicity. The spectral slope of the extinction--free starburst is 
$\beta \simeq -2.8 $ for $t < 0.02$ Gyr , in agrement with what is
generically used \cite{SAGDP99}.
The predicted reddening is $ 0.04 < E (B-V) < 0.4 $ for $t < $0.1 Gyr, in fair
agreement with what is derived in LBGs under the assumption of a universal
$\beta \simeq -2.5 $ \cite{SAGDP99} for the UV slope of galactic spectra without
extinction. The relation of the IRX to $\beta$ reproduces 
the observed trend, provided $f_{\rm H}=$ 100 for the most
extinguished starburts, and points towards an age of
$t=$ 0.01 Gyr for the bursts, 
though part of the scatter can be interpreted as larger ages.

The synthetic SEDs that are produced illustrate the large range of IR/optical 
luminosity ratios. A striking but obvious result is that the optical 
spectrum of a LIRG or ULIRG--type galaxy is extremely sensitive to the 
details of the dust distribution. This makes the predictions on the optical 
counterparts of high-z dusty objects difficult to assess.

We have gathered a small sample of nearby galaxies that have been extensively 
observed at several wavelengths in the optical/IR/submm/radio range. The fit
of the overall continuum with our SEDs provides us with a way of interpolating 
between the data points under assumptions that have a physical meaning. 
We then generate a sequence of galaxies with IR luminosities that span 
several orders of magnitude. 
The spectral trend is very similar to what has 
been shown by e.g. Sanders \& Mirabel \cite*{SM96} on a smaller sample. 
We have established that there is a very good 
correlation between the star formation rate and the bolometric infrared
luminosity, and we have calibrated the relation. In contrast, the UV fluxes in 
these objects are strongly affected by extinction.
Template spectra of the 17 objects listed in table~\ref{t_obj}
generated with {\sc STARDUST}, are available upon
e--mail request ( {\tt devriend@iap.fr} or {\tt guider@iap.fr} ) and can
be used for studies of local and 
distant samples. 

We defer the study of high--z sources to forthcoming papers.
These SEDs have been explicitly designed to be implemented into SAMs 
of galaxy formation and evolution where
the basic free parameters ($t_*$, $t$, and the gas column densities) 
can be computed from general assumptions of the physical processes ruling 
galaxy formation. The remaining uncertainties are due to the possibility that 
dust properties at high z can scale differently with metallicity, that 
perhaps the IMF is not constant, and that gas inflows/outflows can strongly affect galaxy 
evolution.
In a companion paper \cite{DG99}, we will  
implement all this in a simple but explicit cosmological 
framework to try to derive global properties of galaxies and trace
their evolution. 
But the final goal is to work with more realistic codes 
of galaxy formation and evolution, especially those which take into
account the merging histories of dark matter halos 
(e.g. Kauffmann et al. \cite*{KWG93}, Baugh et al. \cite*{BCF96}, 
Somerville and Primack
\cite*{SP99}, Ninin et al. \cite*{NBDG99}).

\begin{acknowledgements}
We are pleased to thank Alessandro Boselli for providing 
us with an electronic version of his data, and J. D. 
thanks Fran\c{c}ois Legrand and 
Michel Fioc for helpful discussions on SSP models. We also acknowledge
stimulating discussions
with Jean--Loup Puget at various stages of this work. The authors are particularly   
grateful to St\'ephane Charlot
for his comments and suggestions that helped improve
this paper, as well as for providing tables of colors and magnitudes
from the GISSEL model. 
\end{acknowledgements}

\end{document}